\documentclass[groupedaddress,
twocolumn,
showkeys, preprintnumbers,
amsmath,amssymb, aps, pre,
draft,
floatfix,]{revtex4-1}

\usepackage{dynlearn}
\usepackage{subfig}
\usepackage{standalone}
\usepackage{amsthm}
\usepackage{mathrsfs}
\usepackage{graphicx}\usepackage{dcolumn}\usepackage{bm}

\theoremstyle{definition}

\usepackage{dynlearn}
\renewcommand{\H}{\operatorname{H}}

\newcommand{\cs}{\causalstate}

\newcommand{\One}{ {\mathbf{1} } }
\newcommand{\MSym}{\MeasSymbol}
\newcommand{\msym}{\meassymbol}

\newcommand{\MxSSet}{\AlternateStateSet}
\newcommand{\MxSMeasure}{\mu}

\newcommand{\MxSDyn}{\mathcal{W}}
\newcommand{\mxst}{\eta}
\newcommand{\mxstalt}{\zeta}
 
\newcommand{\StartMS}{\bra{\pi}}

\newcommand{\simplex}{\Delta}

\newcommand{\BlackwellMeasure}{\mu_B}

\newcommand{\df}{d_0}
\newcommand{\di}{d_1}
\newcommand{\dsc}{d_\mu}

\newcommand{\LCE}{\lambda}
\newcommand{\LCESpectrum}{\Gamma}
\newcommand{\dLCE}{d_\LCESpectrum}

\newcommand{\arXivID}{2102.10487}

\begin{document}

\newcommand{\ourTitle}{Divergent Predictive States:\\
The Statistical Complexity Dimension of\\
Stationary, Ergodic Hidden Markov Processes}

\title{\ourTitle}

\author{Alexandra M. Jurgens}
\email{amjurgens@ucdavis.edu}

\author{James P. Crutchfield}
\email{chaos@ucdavis.edu}
\affiliation{Complexity Sciences Center\\
Physics and Astronomy Department\\
University of California at Davis
One Shields Avenue, Davis, CA 95616}

\date{\today}
\bibliographystyle{unsrt}

\begin{abstract}
Even simply-defined, finite-state generators produce stochastic processes that
require tracking an uncountable infinity of probabilistic features for optimal
prediction. For processes generated by hidden Markov chains the consequences
are dramatic. Their predictive models are generically infinite-state. And,
until recently, one could determine neither their intrinsic randomness nor
structural complexity. The prequel, though, introduced methods to accurately
calculate the Shannon entropy rate (randomness) and to constructively determine
their minimal (though, infinite) set of predictive features. Leveraging this,
we address the complementary challenge of determining how structured hidden
Markov processes are by calculating their statistical complexity
dimension---the information dimension of the minimal set of predictive
features.  This tracks the divergence rate of the minimal memory resources
required to optimally predict a broad class of truly complex processes. 
\end{abstract}

\keywords{Markov process, Lyapunov exponents, fractal dimension, iterated
function system, mixed state, predictive feature, optimal prediction, Blackwell
measure}

\preprint{\arxiv{\arXivID}}

\maketitle

\section{Introduction}
\label{sec:introduction}

A delicate symbiotic relationship lives at the heart of highly complex
systems---the intricate patterns these systems generate arise through an
interplay between determinism and stochasticity. Despite progress identifying
and measuring degrees of randomness and unpredictability, basic questions about
this state of affairs remain. Specifically, how do we quantify ``structure''?
Can we detect a system's emergent patterns and quantify their organization?

Clearly posing these questions and developing the tools to answer them required,
over the recent decades, integrating Turing's computation theory \cite{Turi37a,
Shan56c, Mins67}, Shannon's information theory \cite{Shan48a}, and Kolmogorov's
dynamical systems theory \cite{Kolm56b, Kolm65, Kolm83, Kolm59, Sina59}.
Together they highlighted the central role that \emph{information}---its
generation, transmission, and storage---plays in the organization and
functioning of complex systems.
Drawing from this convergence, \emph{computational mechanics} \cite{Crut12a}
introduced a definition of the structural complexity of stochastic
processes---the \emph{statistical complexity}, which measures the number and
distribution of optimally-predictive features.

Answers to the randomness-structure dichotomy have been carefully detailed and
successfully implemented for stochastic processes that can be optimally
predicted with \emph{countably-many} predictive features
\cite{Crut01a,Marz14e,Shal98a}. However, there are complex systems arising in
many engineering, physical, and biological systems
\cite{Blac57b,Crut89e,Trav11b,Debo12a,Jurg20a} that require an infinite number
of predictive features. Somewhat soberingly, these truly complex systems are
implicated in many natural phenomena, from the geophysics of earthquakes
\cite{Turc97a} and physiological measurements of neural avalanches
\cite{Begg03a} to semantics in natural language \cite{Debo11a} and cascading
failures in power transmission grids \cite{Dobs07a}.

As first established by Blackwell in the 1950s \cite{Blac57b}, calculating the
Shannon entropy rate of even apparently simple processes---such as those
generated by a discrete time, $N$-state hidden Markov chain (HMC)---requires
tracking an uncountably-infinite set of \emph{belief distributions} over the
HMC's states. The following establishes that optimally predicting this class of
processes requires using these belief distributions as predictive features.
These feature sets, living on the $(N-1)$-simplex, are complex: generically
highly ramified and self-similar, they are the support of similarly-complicated
measures \cite{Crut92c}. This renders prediction of even simply-defined
processes very challenging.  Probing their structure is even more difficult.
Nevertheless, the broad popularity and application of HMCs---not only in the
study of complex systems \cite{Crut12a}, but also in coding theory
\cite{Marc11a}, stochastic processes \cite{Ephr02a}, stochastic thermodynamics
\cite{Bech15a}, speech recognition \cite{Rabi86a}, computational biology
\cite{Birney01, Eddy04}, epidemiology \cite{Breto2009}, and finance
\cite{Ryden98}---gives testimony to the ubiquity of truly complex systems, in
both theory and nature.

Our recent work \cite{Jurg20b} addressed this state of affairs, showing how to
produce infinite sets of predictive features and accurately calculate the
entropy rate for processes generated by HMCs. This gave a new and efficient
tool for consistently describing the randomness of truly complex systems.
However, those results did not address the complementary side of the
complex-system paradox---the structural aspect of the interplay between
structure and randomness. Troublingly, for processes with uncountably-infinite
sets of predictive features, the statistical complexity diverges, substantially
circumscribing its usefulness as a metric of system organization. As we will
show, quantifying the structure of these truly complex systems requires a new
approach and new tools and methods.

Historically, the need for such a measure of divergent information storage and
Blackwell's discovery were perhaps anticipated by Shannon's definition in the
1940s of \emph{dimension rate} \cite{Shan48a}:
\begin{align*}
\lambda = \lim_{\delta \to 0} \lim_{\epsilon \to 0} \lim_{T \to \infty}
  \frac{N ( \epsilon, \delta, T)}{T \log \epsilon}
  ~,
\end{align*}
where $N ( \epsilon, \delta, T)$ is the smallest number of elements that may be
chosen such that all elements of a trajectory ensemble generated over time $T$,
apart from a set of measure $\delta$, are within the distance $\epsilon$ of at
least one chosen trajectory. This is the minimal ``number of dimensions''
required to specify a member of a trajectory (or message) ensemble.
Unfortunately, Shannon devotes barely a paragraph to the concept, leaving it
largely unmotivated and uninterpreted. Nonetheless, we take inspiration from
this to develop the \emph{statistical complexity dimension}, the asymptotic
growth rate of the statistical complexity for an uncountably-infinite set of
predictive features. 

We conjecture that the statistical complexity dimension is the same dimension
rate proposed by Shannon. However, the following goes beyond Shannon's brief
mention to provide constructive and accurate methods for determining this
important system invariant. Technically, statistical complexity dimension is
defined as the information dimension of the (self-similar set of) predictive
states. Several distinct steps are involved. Determining this information
dimension requires establishing ergodicity, calculating the Lyapunov spectrum
of an HMC's mixed-state presentation, and applying a suitably modified version
of the Lyapunov-information dimension conjecture from dynamical systems that
connects the spectrum to the dimension.

To highlight the usefulness of these informational quantities, that otherwise
appear rather abstracted from natural systems, it should be noted that the
following and its predecessor \cite{Jurg20b} were proceeded by two companion
articles that applied the theoretical results here to two, rather different,
physical domains. The first analyzed the origin of randomness and structural
complexity engendered by quantum measurement \cite{Vene19a}. The second solved a
longstanding problem on exactly determining the thermodynamic functioning of
Maxwellian demons, aka information engines \cite{Jurg20a}. That is, the
predecessor and the present development (along with a sequel to be announced in
the conclusion) lay out the mathematical and algorithmic tools required to
successfully analyze structure and randomness in these applied problems. Taken
together, we believe the new approach will find even wider use than in these
application areas.

In the following, we introduce a practical and calculable measure of structural
complexity analogous to Shannon's dimension rate in the form of
\emph{statistical complexity dimension} $\dsc$. Section
\ref{sec:HiddenMarkovProcesses} recalls the necessary background in stochastic
processes, hidden Markov chains, and information theory. Section
\ref{sec:MixedStatePresentation} recounts mixed states and their dynamic---the
mixed-state presentation---as well as the connection to iterated function
systems (IFSs) previously demonstrated. The main result follows in Sec.
\ref{sec:StructureInfiniteStateProcesses}, where the Lyapunov-information
dimension conjecture is reviewed and updated to our needs and the statistical
complexity dimension $\dsc$ is introduced. Finally, in Sec. \ref{sec:Example}
$\dsc$ of a three-state parametrized HMC is calculated across a wide region of
parameter space, demonstrating the insights afforded by and computational
efficiency of our methods.

\section{Hidden Markov Processes}
\label{sec:HiddenMarkovProcesses}

Our main objects of study are stochastic processes and the mechanisms that
generate them---hidden Markov chains, mixed states, and the \eM. We touch on
several of their important properties, including stationarity, ergodicity,
randomness, and memory. The reader familiar with the previous work in this
series \cite{Jurg20b} may skip to \cref{sec:ProcessStructure}. 

\subsection{Processes}
\label{sec:Processes}

A \emph{stochastic process} $\Process$ is a probability measure over a
bi-infinite chain $\ldots \, \MSym_{t-2} \, \MSym_{t-1} \, \MSym_{t} \,
\MSym_{t+1} \, \MSym_{t+2} \ldots$ of random variables, each $\MSym_t$ denoted
by a capital letter. A particular \emph{realization} $\ldots \, \msym_{t-2} \,
\msym_{t-1} \, \msym_{t} \, \msym_{t+1} \, \msym_{t+2} \ldots$ is denoted via
lowercase. We assume values $\msym_{t}$ belong to a discrete alphabet
$\MeasAlphabet$. We work with blocks $\MS{t}{t^\prime}$, where the first index
is inclusive and the second exclusive: $\MS{t}{t^\prime} = \MSym_{t} \ldots
\MSym_{t^\prime-1}$. $\Process$'s measure is defined via the collection of
distributions over blocks: $\{ \Pr(\MS{t}{t^\prime}): t < t^\prime, t,t^\prime
\in \mathbb{Z} \}$.

To simplify the development, we restrict to stationary, ergodic processes:
those for which $\Prob(\MS{t}{t+\ell}) = \Prob(\MS{0}{\ell})$ for all $t \in
\mathbb{Z}$, $\ell \in \mathbb{Z}^+$, and for which individual realizations
obey all of those statistics. In such cases, we only need to consider a
process' length-$\ell$ \emph{word distributions} $\Prob(\MS{0}{\ell})$.

A \emph{Markov process} is one that exhibits memory over a single time step:
$\Pr(\MSym_t|\MS{-\infty}{t}) = \Pr(\MSym_t|\MSym_{t-1})$. A \emph{hidden
Markov process} is the output of a memoryless channel \cite{Cove06a} whose
input is a Markov process \cite{Ephr02a}. The central problem that concerns
us arises since the output process generically is not Markov.

\subsection{Presentations}
\label{sec:Presentations}

Directly working with processes---nominally, infinite sets of infinite
sequences and their probabilities---is cumbersome. So, we turn to consider
finitely-specified mechanistic models that generate them.

\begin{Def}
\label{Def:HMC}
A finite-state edge-labeled \emph{hidden Markov chain} (HMC) consists of:
\begin{enumerate}
\setlength{\topsep}{0mm}
\setlength{\itemsep}{0mm}
\item a finite set of \emph{states}
	$\CausalStateSet = \{\causalstate_1, ... , \causalstate_N \}$,
\item a finite alphabet $\MeasAlphabet$ of $k$ \emph{symbols}
	$x \in \MeasAlphabet$, and
\item a set of $N$ by $N$ \emph{symbol-labeled transition matrices}
	$T^{(\msym)}$, $\msym \in \MeasAlphabet$:
	$T^{(\msym)}_{ij} = \Pr(\causalstate_j,\msym|\causalstate_i)$.
\end{enumerate}
\end{Def}

The associated state-to-state transitions are described by the row-stochastic
matrix $T = \sum_{x \in \MeasAlphabet} T^{(x)}$. The \emph{internal-state
Markov chain} is given by $\{\CausalStateSet, T\}$. The asymptotic, stationary
state distribution is $\pi = \{ \Pr(\causalstate), \causalstate \in
\CausalStateSet\}$ and, as a vector, is given by $T$'s left eigenvector
normalized in probability: $\pi = \pi T$.

The stochastic process $\Process$ generated by an HMC is the set of emitted
symbol sequences and their probabilities that arise from specifying an initial
distribution over states $\CausalStateSet$ and following all allowed
state-to-state transitions. $\Process$ is stationary if the initial
distribution is $\pi$.

\begin{figure}
\centering
\includegraphics{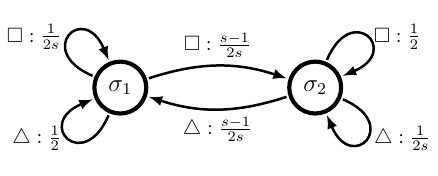}
\caption[text]{A hidden Markov chain with two states
	$\CausalStateSet = \{\sigma_1, \sigma_2 \}$ and two emitted symbols
	$\MeasAlphabet = \{\square, \triangle\}$. It is nonunifilar and
	parametrized with $s \in [1, \infty)$. It becomes unifilar in the limit of
	$s \to \infty$.
	}
\label{Fig:cantormachine}
\end{figure}

A given stochastic process can be generated by any number of HMCs. These
alternative mechanisms are called a process' \emph{presentations}. We now
introduce a structural property of HMCs that has important consequences in
determining a process' randomness and structure.

\begin{Def}
A \emph{unifilar HMC} (uHMC) is an HMC such that for each state
$\causalstate_i \in \CausalStateSet$ and each symbol $\msym \in
\MeasAlphabet$ there is at most one outgoing edge from state
$\causalstate_i$ labeled with symbol $\msym$.
\label{def:UHMC}
\end{Def}

One consequence is that a uHMC's states are \emph{predictive} in the sense that
each is a function of the prior emitted sequence---the past $\ms{-\infty}{t} =
\ldots \msym_{t-2} \msym_{t-1} \msym_t$. Consider an infinitely-long past that,
in the present, has arrived at state $\cs_t$. For uHMCs, it is not required
that this infinitely-long past arrive at a unique state, but it is the case
that any state arrived at by this past must have the same past-conditioned
distribution of future sequences $\Pr(\MS{\infty}{t} | \ms{-\infty}{t})$. We
call this deterministic relationship between the past and the future a
\emph{prediction}.

In comparison, a generative (nonunifilar) HMC may return a set of states with
varying conditional future distributions upon seeing this infinite past. All
that is required for accurate generation is that, if this were to be repeated
many times, averaging over these conditional future distributions returns the
the unique conditional future distribution $\Pr(\MS{\infty}{t} |
\ms{-\infty}{t})$ given by the predictive uHMC state.

Although there can be many presentations for a process $\Process$, there is a
canonical presentation that is unique: a process' \emph{\eM} \cite{Crut12a}.

\begin{Def}
An \emph{\eM} is a uHMC with \emph{probabilistically distinct states}: For each
pair of distinct states $\causalstate_i, \causalstate_j \in \CausalStateSet$
there exists a finite word $w = \ms{0}{\ell-1}$ such that: 
\begin{align*}
\Prob(\MS{0}{\ell} = w|\CausalState_0 = \causalstate_k)
  \not= \Prob(\MS{0}{\ell} = w|\CausalState_0 = \causalstate_j)~.
\end{align*}
\label{def:eM}
\end{Def}

When probabilistically distinct states are enforced for the uHMC, one obtains
the \eM, for which the predictive state itself is unique. This is in contrast
to uHMCs and nonunifilar HMCs, where an infinitely-long past need not lead to
a unique state.

A process' \eM\ is its optimally-predictive, minimal presentation, in the sense
that the set $\CausalStateSet$ of predictive states is minimal compared to all
its other unifilar presentations. That said, $\CausalStateSet$ may be finite,
countably infinite, or uncountably infinite. Since they capture a process'
structure and are not merely predictive, an \eM's states are called \emph{causal
states}.

\subsection{Process Intrinsic Randomness: HMC Entropy Rate}
\label{subsec:EntropyRateHMCs}

A process' intrinsic randomness is the information in the present measurement,
discounted by having observed the preceding infinitely-long history. It is
measured by Shannon's source entropy rate \cite{Shan48a}.

\begin{Def}
\label{Def:EntropyRate}
A process' \emph{entropy rate} $\hmu$ is the asymptotic average
Shannon entropy per symbol \cite{Crut01a}: 
\begin{align}
\hmu = \lim_{\ell \to \infty} \H[\MS{0}{\ell}] / \ell
  ~,
\label{eq:EntropyDensity}
\end{align}
where $H[\MS{0}{\ell}]$ is the \emph{Shannon entropy} of block $\MS{0}{\ell}$:
\begin{align}
H[\MS{0}{\ell}] = - \sum_{\ms{0}{\ell} \in \MeasAlphabet^\ell}
  \Pr(\ms{0}{\ell}) \log_2 \Pr(\ms{0}{\ell})
  ~.
\label{eq:Shannonentropy}
\end{align}
\end{Def}

Given a finite-state unifilar presentation $M_u$ of a process $\Process$, we may
directly calculate the entropy rate from the uHMC's transition matrices
\cite{Shan48a}:
\begin{align}
\hmu = - \sum_{\causalstate \in \CausalState} \pi_\causalstate 
  \sum_{\msym \in \MeasAlphabet}
  \Prob(\msym | \causalstate) \log_2 \Prob(\msym | \causalstate)
  ~.
\label{eq:hmu}
\end{align}
In general, though, for processes generated by nonunifilar HMCs there is no
such closed-form expression for the entropy rate \cite{Blac57b}. For these
processes, the closed-form expression Eq. (\ref{eq:hmu}) applied to the HMC
states and transition matrices substantially misestimates the generated
process' entropy rate.

Addressing this nonunifilar case was the focus of our previous development
\cite{Jurg20b}. We showed that the entropy rate of a general HMC may be
determined using its \emph{mixed states}; reviewed shortly in
\cref{sec:MixedStatePresentation}. Tracking an HMC's mixed states allows one to
find the entropy rate of the generated process and so the latter's intrinsic
randomness.

\subsection{Process Intrinsic Structure}
\label{sec:ProcessStructure}

A process' memory is determined using its \eM. Depending on the specific need,
this may be measured either in terms of the number of causal states
$|\CausalStateSet|$ or the amount of historical Shannon entropy they
store---that is, the \emph{statistical complexity} $\Cmu$.

\begin{Def}
\label{def:Cmu}
A process' \emph{statistical complexity} is the Shannon entropy stored in its
\eM's causal states:
\begin{align}
	\nonumber
\Cmu = & \H[\Pr(\CausalState)] \\
     = & - \sum_{\causalstate \in \CausalState}
	 \pi_\causalstate \log_2 \pi_\causalstate
  ~.
\label{eq:Cmu}
\end{align}
\end{Def}

A process' \eM\ is its smallest uHMC presentation, in the sense that both
|$\CausalStateSet|$ and $\Cmu$ are uniquely minimized by a process' \eM,
compared to all other unifilar presentations. Due to this, the \eM's state
entropy $H[\Pr(\CausalState)]$ is a unique measure of a process'
structural complexity.

A challenge arises similar to that encountered with determining a process'
entropy rate via its nonunifilar HMC presentations: there is no closed-form
expression for the generated process' $\Cmu$. We now turn to give a
constructive answer to this challenge. The preceding presentation types---MC,
HMC, uHMC, and \eM---give a useful path to understanding how a process'
different presentations help or hinder determining process properties. The
strategy in the following turns on yet another presentation type. Here on in,
with nothing else said, reference to an HMC means the general case---a
nonunfilar HMC.

\section{Observer-Process Synchronization}
\label{sec:MixedStatePresentation}

Previously, we introduced mixed-state presentations of HMCs and established
their equivalence to random dynamical systems known as \emph{iterated function
systems} (IFSs) \cite{Jurg20b}. We now briefly review this construction.
(Readers familiar with the previous results may skip to
\cref{sec:StructureInfiniteStateProcesses}.)

Assume that an observer has a finite HMC presentation $M$ for a process
$\Process$. Consider the \emph{observer-process synchronization problem} in
which the observer attempts to determine, at each moment, $M$'s state from
observed data \cite{Jame10a}. Note that this is a dynamical problem---with
every new symbol emitted, the state of $M$ may change. 

Since $M$ is an \emph{hidden} Markov chain, the observer cannot directly detect
the state. With only knowledge of $M$'s structure, the observer's best guess is
that the states occur according to $M$'s internal-state stationary distribution
$\pi$. This distribution describes the proportion of time spent in each state as
observation length approaches infinity, so by making this our initial guess, we
assume $M$ has been generating $\Process$ long enough that we may ignore its
initial condition. The observer then refines this guess by monitoring the output
data $\msym_{0} \, \msym_{1} \, \msym_{2} \ldots$ that $M$ generates, and
comparing this to their knowledge of $M$'s structure. If and when the observer
knows with certainty in which state the process is, they have
\emph{synchronized} to the process.

\subsection{Mixed-State Presentation}
\label{sec:MixedStates}

We formalize the observer-synchronization problem by introducing a new
presentation for a process $\Process$, known as the mixed-state presentation.

\begin{figure*}
\centering
\includegraphics[height=.6\textheight]{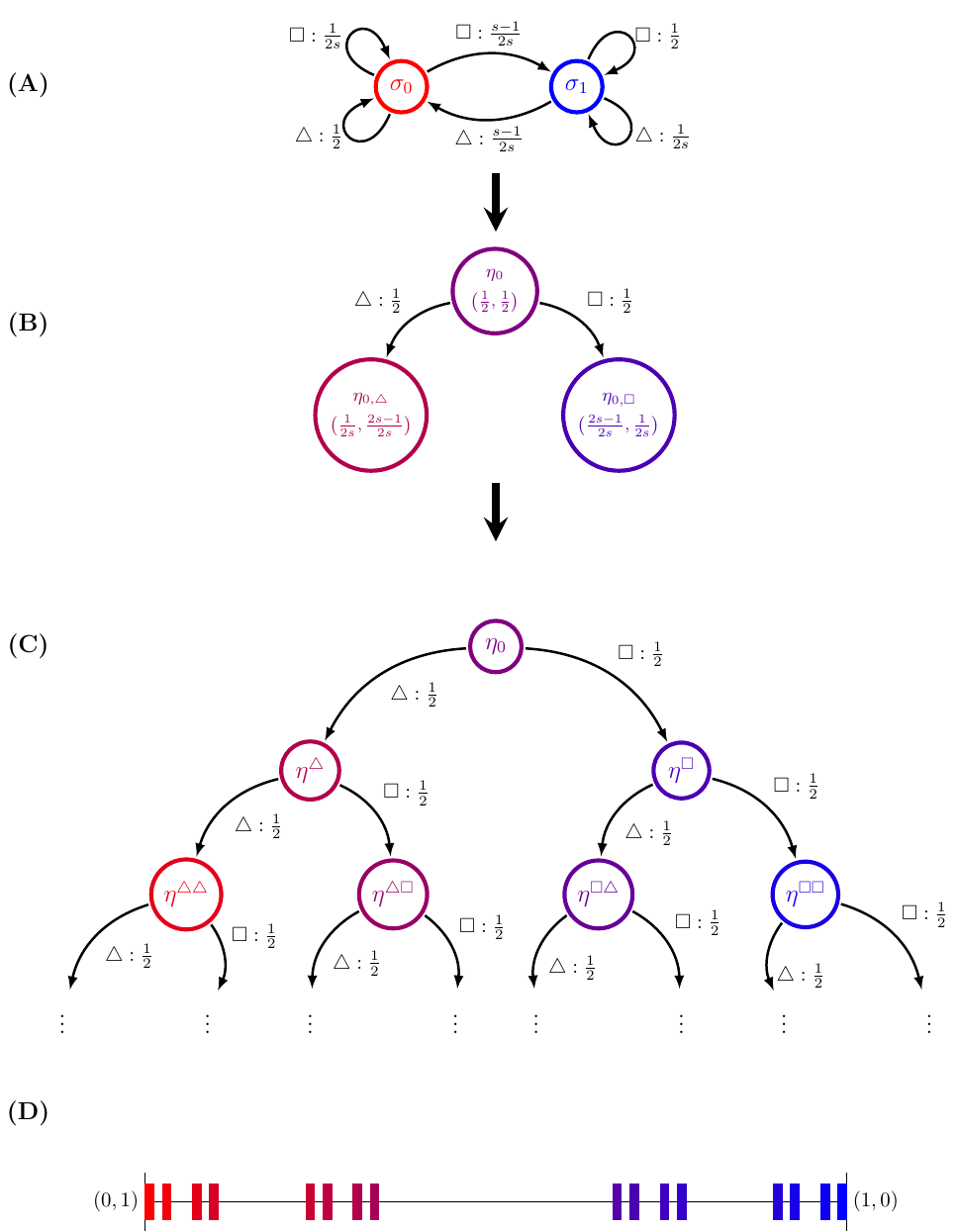}
\caption[text]{Constructing the mixed-state presentation of the $2$-state
	nonunifilar HMC shown in (A): The invariant state-distribution $\pi = (1/2,
	1/2)$. It becomes the first mixed state $\eta_0$ used in (B) to calculate
	the next set of mixed states---those having seen past words of length one.
	(C) In this example, infinitely-many mixed states are generated and
	$\MxSSet$ is the middle-$1/s$ Cantor set. Color is used to indicate the
	relative closeness of each mixed state to the original states in (A). From
	$\eta_0$, one would need to see a word of infinite $\triangle$s to reach
	$\sigma_0$. In (D) the mixed states for $s=3$ are pictured on the
	$1$-simplex---the unit interval from $\eta = (0,1)$ to $\eta = (1,0)$. In
	this representation, the relationship to the Cantor set is visually clear.
	}
\label{Fig:mixedstatemethod}
\end{figure*}

\subsubsection{Mixed-State Set}

Assume we have an $N+1$-state HMC presentation $M$ with $k$ symbols $\msym \in
\MeasAlphabet$. For a length-$\ell$ word $w$ generated by $M$ let $\mxst(w) =
\Prob(\CausalStateSet|w)$ be the observer's \emph{belief distribution} as to the
process' current state after observing $w$: 
\begin{align}
\mxst(w) \equiv
  \Prob(\CausalState_\ell | \MS{0}{\ell}=w, \CausalState_0 \sim \pi)
  ~.
\label{eq:MixedState}
\end{align}
$\MSym \sim \eta$ means that random variable $\MSym$ is distributed according to
$\eta$.

The belief distributions $\mxst(w)$ that an HMC can visit defines its set of
\emph{mixed states}:
\begin{align*}
 \MxSSet = \{ \mxst(w): w \in \MeasAlphabet^+, \Pr(w) > 0 \} ~,
\end{align*}
where $\MeasAlphabet^+$ indicates the set of all words with positive length.
Generically, the mixed-state set $\MxSSet$ for an $N$-state HMC is infinite,
even for finite $N$ \cite{Blac57b}. \Cref{Fig:mixedstatemethod} shows a case
where the HMC generates the Cantor set as its mixed state set. 

When observing a $N$-state machine, the vector $\bra{\mxst(w)}$ lives in the
\emph{(N-1)-simplex} $\simplex^{N-1}$, the set such that:
\begin{align*}
 \{ \eta \in \mathbb{R}^{N} : \langle\eta\ket{\One} = 1, \langle\eta\ket{\delta_i} \geq 0,
  i=1, \dots , N \}
  ~,
\end{align*}
where $\bra{\delta_i} = \begin{pmatrix} 0 & 0 & \dots & 1 & \dots & 0
\end{pmatrix}$ and $\ket{\One} = \begin{pmatrix} 1 & 1 & \dots & 1
\end{pmatrix}$. We use this notation for components of the mixed-state
vector $\mxst$ to avoid confusion with temporal indexing.

\subsubsection{Mixed-State Dynamic}

The set of mixed states itself does not constitute a presentation. We must
augment it with the state dynamic, which gives the transition probabilities
between mixed states. Let the initial mixed state be the invariant probability
$\pi$ over $M$'s states, so $\bra{\mxst_0} = \StartMS$. In the context of the
mixed-state dynamic, mixed-state subscripts denote time. The probability of
transitioning from $\bra{\mxst(w)}$ to $\bra{\mxst(w\msym)}$ on observing
symbol $\msym$ follows from \cref{eq:MixedState} immediately; we
have:\begin{align*}
\Pr(\mxst(w\msym) | \mxst(w)) = \Pr(\msym|\CausalState_\ell \sim \mxst(w))
 ~.
\end{align*}
This defines the mixed-state transition dynamic $\MxSDyn$.

\begin{figure*}
\centering
\subfloat[Mixed state attractor for 3-state ``alpha'' machine. 
  \label{fig:simplex_example_alpha}]{
      \includegraphics[width=0.32\textwidth]{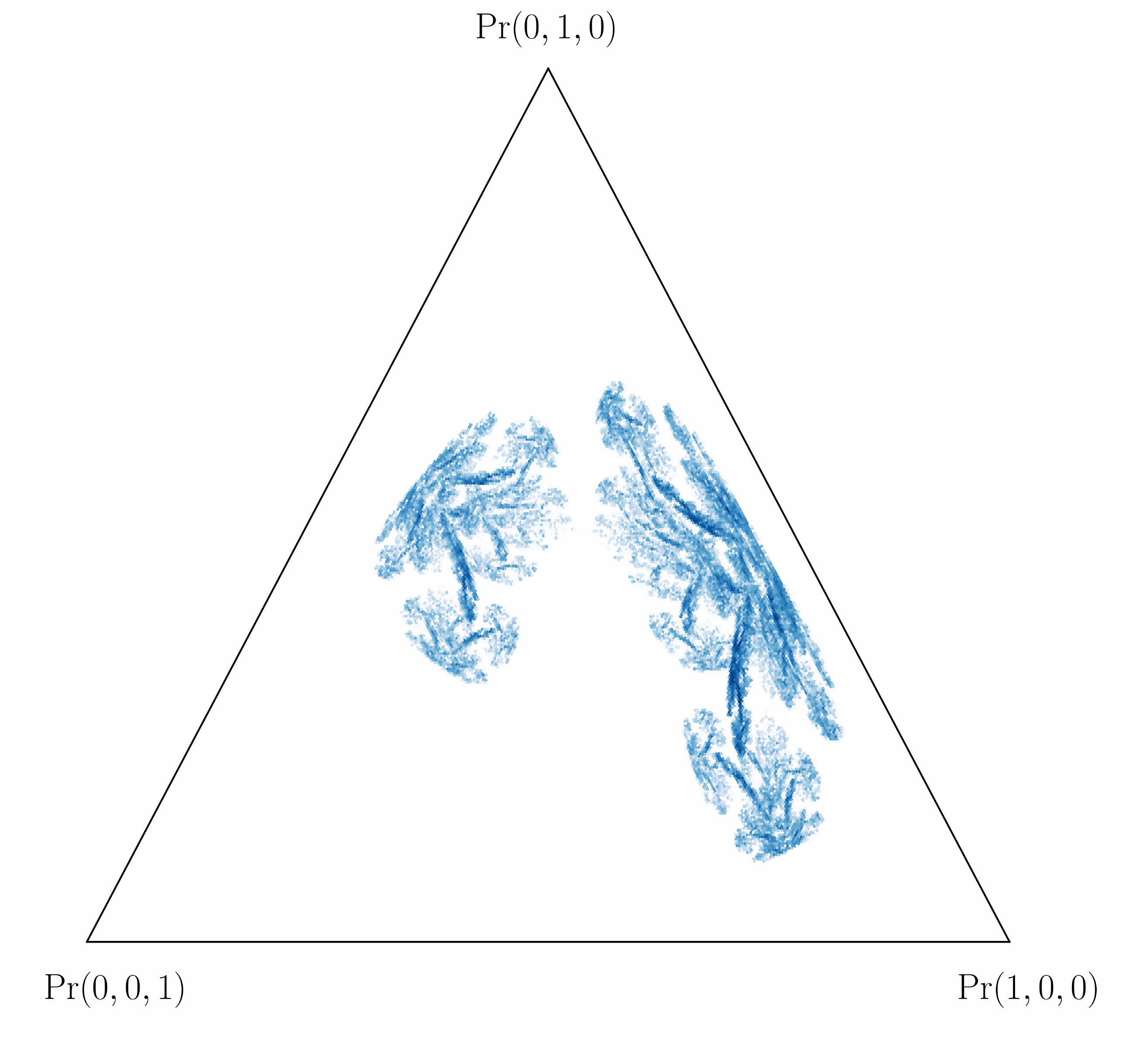}
  } 
\subfloat[Mixed state attractor for 3-state machine from
\cref{eq:sarah_machine}, with $\alpha = 0.6$ and $x = 0.1$. 
\label{fig:simplex_example_sarah}]{
      \includegraphics[width=0.32\textwidth]{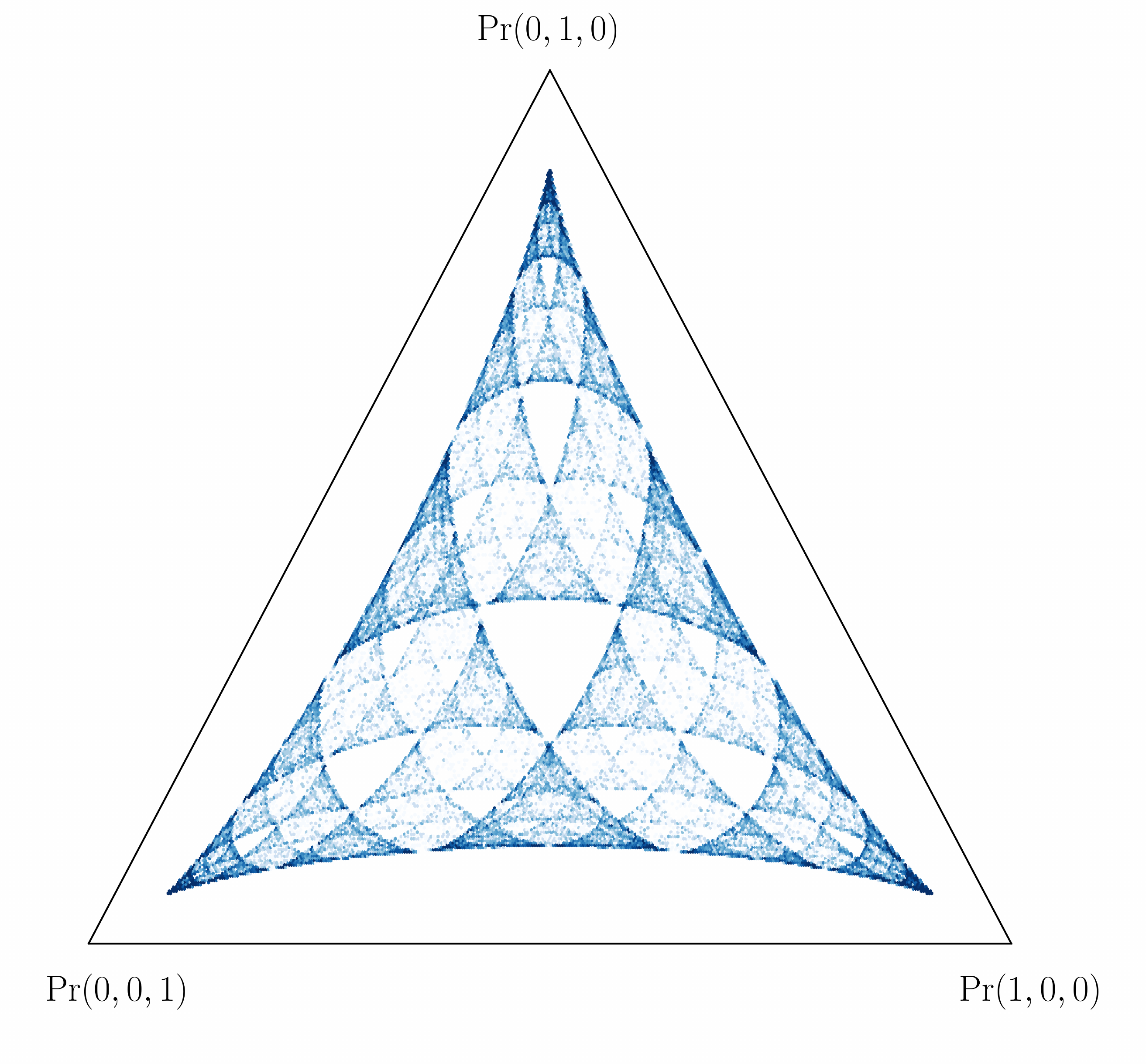}
  }
\subfloat[Mixed state attractor for 3-state ``beta'' machine.
\label{fig:simplex_example_beta}]{
      \includegraphics[width=0.32\textwidth]{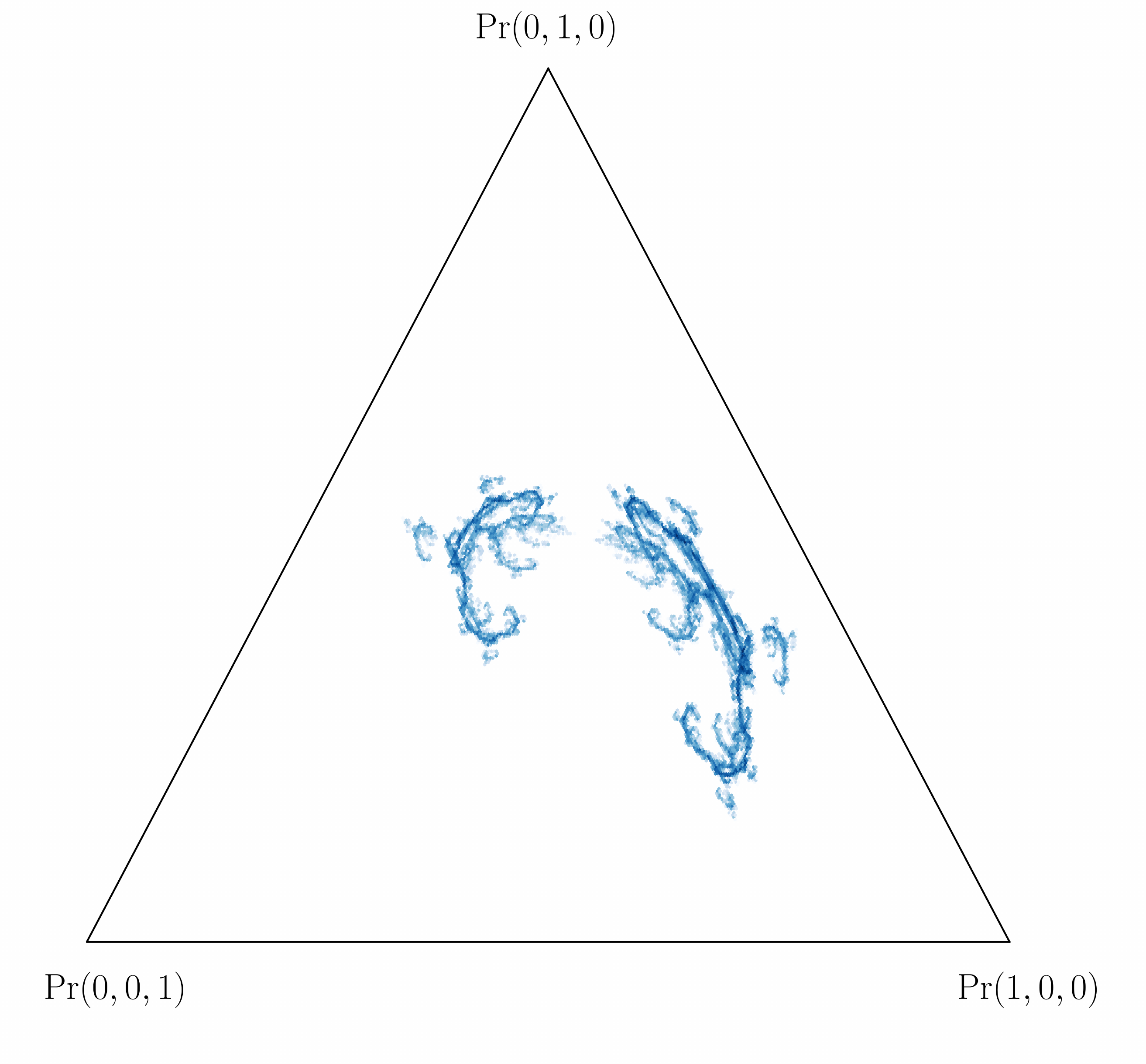}
  }
\caption{Simple HMCs generate MSPs with a wide variety of structures,
	many fractal in nature. Each subplot displays $10^4$ mixed states of a
	different, highly-nonunifilar $3$-state hidden Markov chain. The HMCs
	themselves are specified in \cref{app:nonunifilarHMCs}.
	}
\label{fig:simplex_examples}
\end{figure*}
 
Given $M$'s symbol-labeled transition matrices, we can specify the mixed-state
dynamic $\MxSDyn$ in closed form. First, the probability of generating symbol
$\msym$ when in mixed state $\mxst$ is:
\begin{align}
  \Pr(\msym|\mxst) = \bra{\mxst} T^{(\msym)} \ket{\One}
  ~,
\label{eq:SymbolFromMixedState}
\end{align}
where $T^{(\msym)}$ is the symbol-labeled transition matrix associated with the
symbol $\msym$. Now, given a mixed state at time $t$, we may calculate the
probability of seeing each $\msym \in \MeasAlphabet$. Upon seeing symbol
$\msym$, the current mixed state $\bra{\mxst_t}$ is updated according to:
\begin{align}
  \bra{\mxst_{t+1, \msym}}
  = \frac{\bra{\mxst_t} T^{(\msym)}} {\bra{\mxst_t} T^{(\msym)} \ket{\One}}
  ~.
\label{eq:MxStUpdate}
\end{align}

Equation (\ref{eq:MxStUpdate}) tells us that, by construction, the MSP is
unifilar, since each possible output symbol uniquely determines the next (mixed)
state. Taken together, \cref{eq:SymbolFromMixedState,eq:MxStUpdate} define the
mixed-state transition dynamic $\MxSDyn$ as:
\begin{align*}
\Pr(\mxst_{t+1},\msym|\mxst_t) & = \Pr(\msym|\mxst_t) \\
                               & = \bra{\mxst_t} T^{(\msym)} \ket{\One} 
  ~,
\end{align*}
for all $\mxst \in \MxSSet$, $\msym \in \MeasAlphabet$.

Together the mixed states and their dynamic define an HMC that is unifilar by
construction. This is a process' \emph{mixed-state presentation} (MSP)
$\MSP(\Process) = \{\MxSSet, \MxSDyn \}$.

\subsection{Constructing the Mixed-State Presentation}
\label{sec:MixedStateMethod}

To explicitly find the MSP $\MSP = \{\MxSSet, \MxSDyn \}$ for a given HMC $M$ we
apply \emph{mixed-state construction}:

\begin{enumerate}
\setlength{\topsep}{-3mm}
\setlength{\itemsep}{-1mm}
\item Set $\MSP = \{ \MxSSet = \emptyset, \MxSDyn = \emptyset \}$.
\item Calculate $M$'s invariant state distribution: $\pi = \pi T$.
\item Take $\mxst_0$ to be $\StartMS$ and add it to $\MxSSet$.
\item For each current mixed state $\mxst_t \in \MxSSet$, use
	\cref{eq:SymbolFromMixedState} to calculate $\Pr(\msym | \mxst_t )$ for
	each $\msym \in \MeasAlphabet$. 
\item For $\mxst_t \in \MxSSet$, use \cref{eq:MxStUpdate} to find the updated
	mixed state $\mxst_{t+1, \msym}$ for each $\msym \in \MeasAlphabet$ with
	$\Pr(\msym|\eta_t) >0 $. 
\item Add $\mxst_{t}$'s transitions to $\MxSDyn$ and each $\mxst_{t+1, x}$
	to $\MxSSet$. 
\item For each new $\mxst_{t+1}$, repeat steps 4-6 until no new mixed states
	are produced.
 \end{enumerate}
This algorithm need not terminate, as shown in \cref{Fig:mixedstatemethod},
which depicts the MSP construction for the HMC in \cref{Fig:cantormachine}.
However, it can terminate for HMCs described by finite-state \eMs. When working
with HMCs for which the algorithm does not terminate, one must impose a limit on
the number of generated mixed-states, effectively setting a level of
approximation for $\MxSSet$.

One may ask, given that mixed-state construction returns a unifilar HMC of the
underlying process, is the MSP the same as the \eM? It is not guaranteed to be
so, as indeed is the case in \cref{Fig:mixedstatemethod}. While the
\emph{Cantor machine} there generates an uncountably-infinite set of mixed
states, the underlying process' \eM is a single-state fair coin---which has a
one-state HMC as its \eM. We see this by noting that the symbol-branching
probabilities depicted in \cref{Fig:mixedstatemethod}(C) are identical for
every generated mixed state. This may seem a cause for concern. Perhaps we
overestimated the necessary state size for the process in
\cref{Fig:mixedstatemethod} by an infinite factor. However, this case is both
rare and easy to check for, as discussed in \cref{app:MinimalityofMSP}. By
applying a simple check on the uniqueness of mixed states as they are
constructed, we can confirm if the MSP is the underlying process' \eM. Unless
otherwise noted, we take this to hold.

Although the most common purpose of applying mixed-state construction is to
\emph{unifilarize} an HMC, we may find the MSP of uHMCs as well. The MSPs of
unifilar presentations are interesting and contain additional information beyond
the initial unifilar presentation. For example, they typically contain transient
causal states and these are employed to calculate many complexity measures that
track convergence statistics \cite{Crut13a}. The following only considers
nontransient (asymptotically recurrent) mixed states.

The following also narrows the focus to mixed-state presentations of
nonunifilar HMCs, which typically have an infinite mixed-state set $\MxSSet$.
By way of emphasizing a principle result---and Blackwell's original point
\cite{Blac57b}---applying mixed-state construction even to nominally simple,
finite-state, finite-alphabet nonunfilar HMCs results in an explosion of mixed
states.  \Cref{fig:simplex_examples} gives three examples of MSPs with fractal
mixed-state sets $\MxSSet$, each generated by a three-state nonunifilar HMC.

\subsection{MSP as an IFS}
\label{sec:MixedStateasanIFS}

Specifying MSP construction in this way reveals that generating mixed states is
a type of random dynamical system known as \emph{place-dependent iterated
function system} (IFS) \cite{Jurg20b}. For finite $k$ and space $\simplex$,
place-dependent IFSs are characterized by a set of \emph{mapping functions}:
\begin{align*}
  \left\{ f^{(x)} : \simplex \to \simplex \mid x \in \{ 0, 1 \dots k \} \right\}
  ~,
\end{align*}
and associated \emph{probability functions}:
\begin{align*}
  \left\{ p^{(x)} : \simplex \to \left[ 0,1 \right] \mid x \in \{ 0, 1 \dots k\}  \right\}
  ~.
\end{align*}

A place-dependent IFS generates a stochastic process over $\eta \in
\simplex^N$ as follows: Given an initial position $\mxst_0 \in \simplex^N$, the
probability distribution $\{ p^{(\msym)}(\mxst_0) : \msym = 1, \dots, k \}$ is
sampled. According to the sample $\msym$, apply $f^{(\msym)}$ to map $\mxst_0$
to the next position $\mxst_1 = f^{(\msym)}(\mxst_0)$. Resample $x$ from the
distribution and continue, generating $\mxst_0, \mxst_1, \mxst_2, \ldots$.

An $N$-state HMC's associated mixed-state presentation defines a
place-dependent IFS over the $N-1$ simplex, with each symbol-labeled transition
matrix $T^{(x)}$ defining a mapping function (\cref{eq:MxStUpdate}) and
associated probability function (\cref{eq:SymbolFromMixedState}). Our previous
results showed that an IFS defined by an ergodic HMC generates an ergodic
mixed-state process and has a unique attractor---the set of mixed states
$\MxSSet$ \cite{Jurg20b}. Additionally, this attractor has a unique,
attracting, invariant measure known as the \emph{Blackwell measure}
$\BlackwellMeasure(\MxSSet)$.

\section{Structure of Infinite-State Processes}
\label{sec:StructureInfiniteStateProcesses}

Our prior development showed how to use the MSP to find a process' intrinsic
randomness---in the form of Shannon entropy rate $\hmu$ \cite{Jurg20b}. Our
goal here is to complement the measure of randomness with a measure of
structure.

Recall that for processes generated by finite unifilar HMCs a unique minimal
presentation---the $\epsilon$-machine---exists. Due to the latter's minimality,
for such processes we can then define the statistical complexity $\Cmu$ using
\cref{eq:Cmu}. In contrast, general (nonunifilar) HMCs have no such canonical minimal presentation. And so, the Shannon entropy of their presentation states
is ambiguous. However, \cref{sec:MixedStatePresentation} demonstrated how to
find the mixed-state presentation $\MSP(M)$ of a nonunifilar HMC $M$, which
operation unifilarizes it. A heavy cost is levied---explosion of the
mixed-state set. Nonetheless, using the MSP to develop a measure of structure
for processes generated by HMCs is naturally appealing, since the MSP is
unifilar and unique \cite{Jurg20b}. Indeed, in most cases, the MSP is the \eM
of the underlying process and so minimal (see \cref{app:MinimalityofMSP}). And,
when not, we can minimize the MSP by merging equivalent mixed states as we
construct them.

However, the naive approach of simply measuring structure with statistical
complexity $\Cmu$ introduces a problem: the statistical complexity diverges for
an HMC with an uncountably-infinite state set $\MxSSet$. In general, MSPs of
HMCs are uncountably-infinite state, precluding distinguishing them via $\Cmu$.
This being said, it is clear visually from \cref{fig:simplex_examples} that
HMCs with uncountably-infinite state spaces still have significant and distinct
structures. We wish to find a way to measure and distinguish such structure.
For this, we take inspiration from Shannon's dimension rate \cite{Shan48a} and
call on a familiar tool.

Fractal dimension measures the rate at which a chosen size metric of a set
diverges with the scale at which the set is observed
\cite{Reny59a,Mand82a,Edga90a,Falc90a,Pesi97a}. Fractal dimension is also useful
to probe the ``size'' of objects when cardinality is not informative. For
example, the mixed-state presentation, generically, has an uncountable infinity
of causal states. That observation is far too coarse, though, to distinguish the
clearly distinct mixed-state sets $\MxSSet$ in Fig. \ref{fig:simplex_examples}.
Each is uncountably infinite, but the $\MxSSet$'s geometries differ. Determining
their fractal and other dimensions will allow us to distinguish them and allow
us to introduce additional insights into the original process' intrinsic
information processing.

\subsection{Dimensions}
\label{sec:FractalDimensions}

Consider the mixed-state set $\MxSSet$ on the simplex for an $N$-state HMC $M$
that generates a process $\Process$. We consider two types of dimension for
$\MxSSet$: the Minkowski-Bouligand or box-counting dimension, often simply
called the fractal dimension, and the information dimension.

To calculate the first, coarse-grain the $N$-simplex with evenly spaced
subsimplex cells of side length $\epsilon$. Let $\mathcal{F}(\epsilon)$ be the
set of cells that encompass at least one mixed state. Then $\MxSSet$'s
\emph{box-counting dimension} is:
\begin{align}
\df ( \MxSSet ) = - \lim_{\epsilon \to 0} 
  \frac{\log | \mathcal{F(\epsilon)} |}{ \log \epsilon}
  ~,
\label{eq:BoxCountingDimension}
\end{align}
where $| C |$ is the size of set $C$.

The information dimension tracks how the Blackwell measure
$\BlackwellMeasure(\MxSSet)$ scales with $\epsilon$. Let each cell in
$\mathcal{F}(\epsilon)$ be a state and approximate the dynamic over $\MSP(M)$
by grouping all transitions to and from states encompassed by the same cell.
This results in a finite-state Markov chain that generates an approximation of
the original mixed-state process and has a stationary distribution
$\mu(\mathcal{F}(\epsilon))$. Then $\BlackwellMeasure(\MxSSet)$'s
\emph{information dimension} is:
\begin{align}
\di (\BlackwellMeasure(\MxSSet) )
  & =  \lim_{\epsilon \to 0} 
  \frac{ H_{\MxSMeasure} [\mathcal{F}(\epsilon)] }{ \log \epsilon }
  ~,
\label{eq:InformationDimension}
\end{align}
where $H_{\MxSMeasure} [\mathcal{F}(\epsilon)] = - \sum_{C_i \in
\mathcal{F}(\epsilon)} \MxSMeasure(C_i) \log \MxSMeasure (C_i)$ is the Shannon
entropy over the set $\mathcal{F}(\epsilon)$ of cells that cover attractor
$\MxSSet$ with respect to $\MxSMeasure$.

\subsection{Dimensions and Scaling of HMCs}
\label{sec:DimensionsofHMCs}

These dimensions give two complementary resource-scaling laws for HMC-generated
processes. Rearranging \cref{eq:BoxCountingDimension}, we see that the number
of mixed states in our finite-state approximation to $\MSP(M)$ scales
algebraically with $\MxSSet$'s box-counting dimension:
\begin{align}
    |\mathcal{F} (\epsilon) | \sim \epsilon^{- \df (\MxSSet)}
	~.
\label{eq:StateScalingRelationship}
\end{align}
In other words, for an uncountably infinite MSP, the exponential growth rate of
mixed states is $\df (\MxSSet)$.

Similarly, the entropy of the Blackwell measure scales with the information
dimension. Rearranging \cref{eq:InformationDimension} shows that the state
entropy of the finite-state approximation to $\MSP(M)$ scales logarithmically
with $\MxSSet$'s information dimension with respect to the Blackwell measure:
\begin{align}
    H_\mu [\mathcal{F}] \sim \di (\BlackwellMeasure) \cdot \log \epsilon
	~.
\label{eq:StateEntropyScalingRelationship}
\end{align}
As $\epsilon \to 0$, $|\mathcal{F}|$ and $H_{\mu}(\mathcal{F})$ diverge and
$\df$ and  $\di$ are the divergence rates, respectively. The remainder
focuses on $\di$ as applied to the \eM, for which it describes the rate of
divergence of statistical complexity $\Cmu$.

\section{Statistical Complexity Dimension}
\label{sec:StatisticalComplexityDimension}

We refer to the information dimension $\di(\mu)$ of the \eM the
\emph{statistical complexity dimension} $\dsc$.

Applying $\di$ to the Blackwell measure $\BlackwellMeasure(\MxSSet)$ gives the
rate of divergence of $\Cmu$ as one constructs increasingly better finite-state
approximations to the infinite-state \eM. In this way, $\dsc$ describes the
divergence of memory resources when attempting to optimally predict a process
that requires an uncountably-infinite number of predictive features. This is a
unique, minimal description of the process' structural complexity. This solves
the challenge posed in the introduction: quantifying structure for a broad class
of truly complex systems. 

When a process may be optimally predicted with a finite number of predictive
features, the statistical complexity dimension vanishes. In this case, the more
relevant complexity measure is the original \eM statistical complexity $\Cmu$,
which is is finite. When a process requires uncountably-infinite causal states,
but may be generated with a minimal $N$-state (nonunifilar) HMC, the statistical
complexity is less than or equal to $N-1$. This is the associated IFS's
\emph{embedding dimension}, since the mixed states lie in a space of dimension
$N-1$. 

Unfortunately, directly calculating the information dimension using
\cref{eq:InformationDimension}---and therefore calculating the statistical
complexity dimension $\dsc$---is nontrivial, as it requires estimating a fractal
measure. Fortunately, to calculate the $\dsc$ of the mixed state attractor
$\MxSSet$, we can leverage the associated generating dynamical system (see
\cref{sec:MixedStateasanIFS}).

\subsection{Dimension from Dynamical (In)Stabilities}
\label{sec:LyapunovDimension}

We can link the information dimension of an MSP's mixed state set $\MxSSet$ to
the stability properties of the associated IFS. This starts with determining the
local time-average stability and instability of orbits within an attractor via
the \emph{spectrum of Lyapunov characteristic exponents} $\LCESpectrum =
\{\LCE_1, \ldots, \LCE_N : \LCE_i \geq \LCE_{i+1} \}$ \cite{Shim79a,Bene80a}.
Individual LCEs $\LCE_i$ measure the average local growth or decay rate of orbit
perturbations. The net result is a list of quantities that indicate long-term
orbit instability ($\LCE_i > 0)$ and orbit stability ($\LCE_i < 0)$ in
complementary directions. Usefully, their sum gives the net state-space
divergence---volume loss for dissipative systems. The sum of the positive LCEs
is the dynamical system's entropy rate
\cite{Pesi76a,Ruel79a,Bene78a,Shim79a,Pesi97a}---the net information generation.

To motivate our present use of the Lyapunov spectrum, it will help to develop a
simple intuition for how the LCEs relate to dimension. Both of our previously
defined dimensional quantities---\cref{eq:InformationDimension} and
\cref{eq:BoxCountingDimension}---depended on the growth rate of cells needed to
cover the attractor as we take the side length of the cells to zero. First,
imagine the attractor of a two-dimensional map with LCEs $\LCE_1 > 0 > \LCE_2$
and whose state space is covered with equally spaced squares of side length
$\epsilon$. After iterating the map $q$ times, for $\epsilon$ small enough, the
local action of the map is approximately linear. From the LCE definition, this
means it takes the initial square cells to rectangles of average length
$(e^{\LCE_1 q})\epsilon$ and average width $(e^{\LCE_2 q})\epsilon$. Now,
consider covering the attractor with a new set of squares of side length
$(e^{\LCE_2 q})\epsilon$. We can see by inspection that this requires roughly
$e^{\left(\LCE_1 / \LCE_2 \right) q}$ squares per rectangle. In this way,
the Lyapunov exponents relate to the scalings measured by dimension
quantities \cite{Farm83}.

Indeed, this relationship has resulted in the definition of a \emph{Lyapunov
dimension} $\dLCE$ in terms of the spectrum $\LCESpectrum$ \cite{Kapl79a}:
\begin{align}
\dLCE = \left\{
  \begin{aligned}
	& k + \frac{\sum_i^k \LCE_i}{| \LCE_{k+1} | } ~, & \sum_i^N \LCE_i < 0 \\
    & N ~, & \sum_i^N \LCE_i \geq 0
  \end{aligned}
  \right.
  ~,
\label{eq:LyapunovDimension}
\end{align}
where $k$ is the largest index such that the summation $\sum_i^k \LCE_i$
remains positive. If $\LCE_1 < 0$, $\dLCE = 0$. Its name helps distinguish the
conditions under which the relationships between the various dimensions actually
hold. (There are many conditions and system classes that this summary
necessarily leaves out.)

It has been shown that for any ergodic, invariant probability measure
$\MxSMeasure$:
\begin{align*}
    \di (\MxSMeasure) \leq \dLCE ~,
\end{align*}
with equality when $\MxSMeasure$ is a Sinai-Bowen-Ruelle (SBR) measure
\cite{Ledrap85}. Furthermore, it was conjectured that for ``typical dynamical
systems'', the Lyapunov dimension $\dLCE$ equals the information dimension
$\di$ \cite{Kapl79a}. This remarkable relationship directly relates a system's
dynamics to the geometry and natural measure of its attractor. Additionally,
and usefully, \cref{eq:LyapunovDimension} gives us a tractable method to find
the information dimension of an attractor generated by a dynamical system when
the equality holds and an upper bound when it does not.

\begin{figure*}
\centering
\includegraphics[width=.9\textwidth]{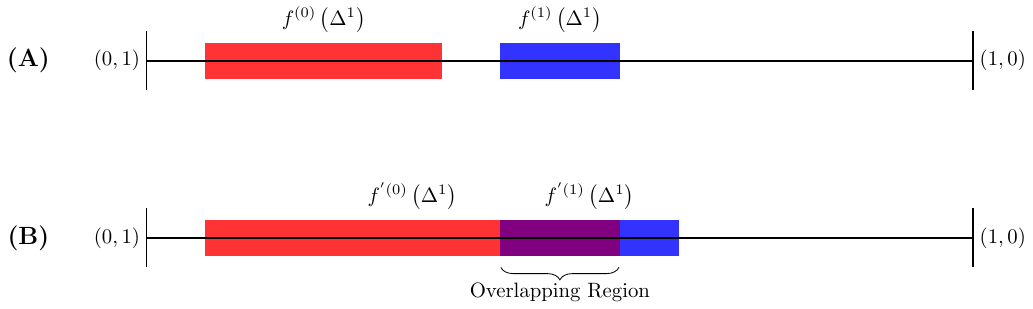}
\caption[text]{Overlap problem on the $1$-simplex $\simplex^1$: Two distinct
	IFSs are considered, each with two mapping functions. The images of the
	mapping functions over the entire simplex are depicted in red and blue.
	(A) Images of the mapping functions $f^{(0)}$ and $f^{(1)}$ do not
	overlap---every mixed state $\eta_t \in \MxSSet$ has a unique pre-image.
	(B) Images of the mapping functions overlap (purple Overlapping
	Region)---there exist $\eta_1, \eta_2 \in \MxSSet$ such that
	$f^{'(0)}(\eta_1) = f^{'(1)}(\eta_2) = \eta_3$. This case is an
	\emph{overlapping IFS}.
  }
\label{Fig:simplex_overlap}
\end{figure*}

\subsection{Calculating Statistical Complexity Dimension}
\label{sec:CalculatingInformationDimension}

We now have in hand two important pieces. First, the definition of statistical
complexity dimension $\dsc$ as the information dimension of an \eM. Second, we
have a bound on the information dimension of an attractor, given knowledge of
the generating system's dynamics. To complete our picture, we now address the
final puzzle piece.

As \cref{sec:LyapunovDimension} discussed, there is a direct relationship
between the information dimension of a chaotic attractor and the dynamics of the
system to which the attractor belongs. Furthermore, as discussed in
\cref{sec:MixedStateasanIFS}, every HMC has an associated random dynamical
system---the iterated function system (IFS)---which has the HMC's set of mixed
states $\MxSSet$ as its unique attractor. Combining these two facts allows us to
exactly calculate $\dsc$ in many cases of interest. 

An advantage of working with IFSs defined by HMCs is a clean division exhibited
by their Lyapunov spectrum. It has been shown that the entropy rate of the
generated process is equivalent to the largest Lyapunov exponent
\cite{Jacquet08,Holliday06}. Calculating this was the main topic of the prequel
to the present work \cite{Jurg20b}. Furthermore, due to the contractivity of
the IFS mapping functions, all other Lyapunov exponents will necessarily be
negative. For a review of calculating the Lyapunov exponents for IFSs,
see \cref{app:LyapunovExponents}.

Therefore, the Lyapunov dimension of an IFS is:
\begin{align}
\widetilde{\dLCE} = \left\{
  \begin{aligned}
	& k - 1 + \frac{\hmu + \sum_i^{k-1} \lambda_i}{|\lambda_{k}|} ,
	& \hmu + \sum_i^N \lambda_i < 0 \\
	& N ~,
	& \hmu + \sum_i^N \lambda_i \geq 0
  \end{aligned}
  \right.
  ,
\label{eq:IFSLyapunovDimension}
\end{align}
where $k$ is now the largest index for which $\hmu + \sum_i^k \lambda_i > 0$.
This is simply \cref{eq:LyapunovDimension}, as if $\hmu$ was the largest Lyapunov
exponent. 

Under specific technical conditions to be discussed shortly, the IFS $\dLCE$ is exactly equal to the information dimension of the IFS's attractor
and, therefore, is $\dsc$ \cite{Baran15b}. In
general, relaxing those conditions, $\widetilde{\dLCE}$ upper bounds the
statistical complexity dimension:
\begin{align}
  \widetilde{\dLCE} \geq \dsc
  ~. 
\label{eq:DSCbound}
\end{align}

Assembling these pieces together determines the basic algorithm to calculate
(or bound) the statistical complexity dimension:
\begin{enumerate}
  \setlength{\topsep}{-3mm}
  \setlength{\itemsep}{-1mm}
  \item For an $N$-state HMC $M$ with $|\MeasAlphabet | = k$, write down the
  associated IFS with $k$ symbol-labeled mapping functions and probability
  functions. 
  \item Calculate the entropy rate $\hmu$ using the Blackwell limit (see
  \cite{Jurg20b}).
  \item Calculate the negative Lyapunov exponents $\left\{ \LCE_1, \dots, \LCE_{N-1}
  \right\}$ (see \cref{app:LyapunovExponents}).
  \item Compute the Lyapunov dimension $\dLCE$ using \cref{eq:IFSLyapunovDimension}.
\end{enumerate}
As mentioned, in specific cases, the Lyapunov dimension is exactly equal to the
statistical complexity dimension, and our task is complete. However, there are
major technical concerns with when we have only the bound in \cref{eq:DSCbound}
and with its tightness then.

\subsection{The Overlap Problem}
\label{sec:OverlapProblem}

A subtle disadvantage of working with IFSs is a direct result of the
stochastic nature of them as random dynamical systems. We must consider the
\emph{overlap problem}, which concerns the ranges of the symbol-labeled mapping
functions $f^{(i)}$, illustrated in \cref{Fig:simplex_overlap}. Specifically,
the problem means that we must distinguish between IFSs that meet the open set
condition and those that do not.

\begin{Def}
\label{Def:OSC} 
An iterated function system with mapping functions $f^{(\mxst)} : \simplex^N
\to \simplex^N$ satisfies the \emph{open set condition} (OSC) if there exists
an open set $U \in \simplex^N$ such that for all $\mxst, \mxstalt \in
\simplex^N$:
\begin{align*}
  f^{\mxst}(U) \cap f^{\mxstalt}(U) = \emptyset, \:\:\:\: \mxst \ne \mxstalt
  ~. 
\end{align*}
IFSs that meet the OSC are \emph{nonoverlapping IFSs}.
\end{Def}

When the images of the symbol-labeled mappings overlap the inequality in
\cref{eq:DSCbound} is strict. To briefly outline the consequences, for an
overlapping IFS the entropy rate $\hmu$ does not accurately capture state-space
expansion. And, this causes the IFS $\dLCE$ (\cref{eq:IFSLyapunovDimension}) to
overestimate the information dimension. As a rule of thumb, the degree to which
the mappings overlap determines the magnitude of the bound's error. The impact
of overlaps is significant. It is explored both in \cref{sec:Example}, where we
calculate the statistical complexity dimension for HMCs with and without
overlap, as well as in the sequel, which diagnoses the problem's origins and
provides an algorithmic solution.

For now, to give a workable approach, we simply introduce two extra steps to
the $\dsc$ algorithm from the previous section:
\begin{enumerate}
  \setlength{\topsep}{-3mm}
  \setlength{\itemsep}{-1mm}
  \setcounter{enumi}{4}
  \item Determine if the Open Set Condition is met using the mapping functions
  $f^{(i)}$. 
  \item If the OSC is not met, estimate the degree of overlap to determine the
  closeness of the bound on $\dsc$. 
\end{enumerate}

\subsection{Statistical Complexity Dimension for Processes Generated by Two-State HMCs}
\label{sec:TwoStateMachines}

Finally, we analyze two-state HMCs, for which \cref{eq:IFSLyapunovDimension}
simplifies significantly. When there is no overlap in the maps, we have exact
results for $\dsc$.

\begin{figure*}
\centering
\includegraphics[width=\textwidth]{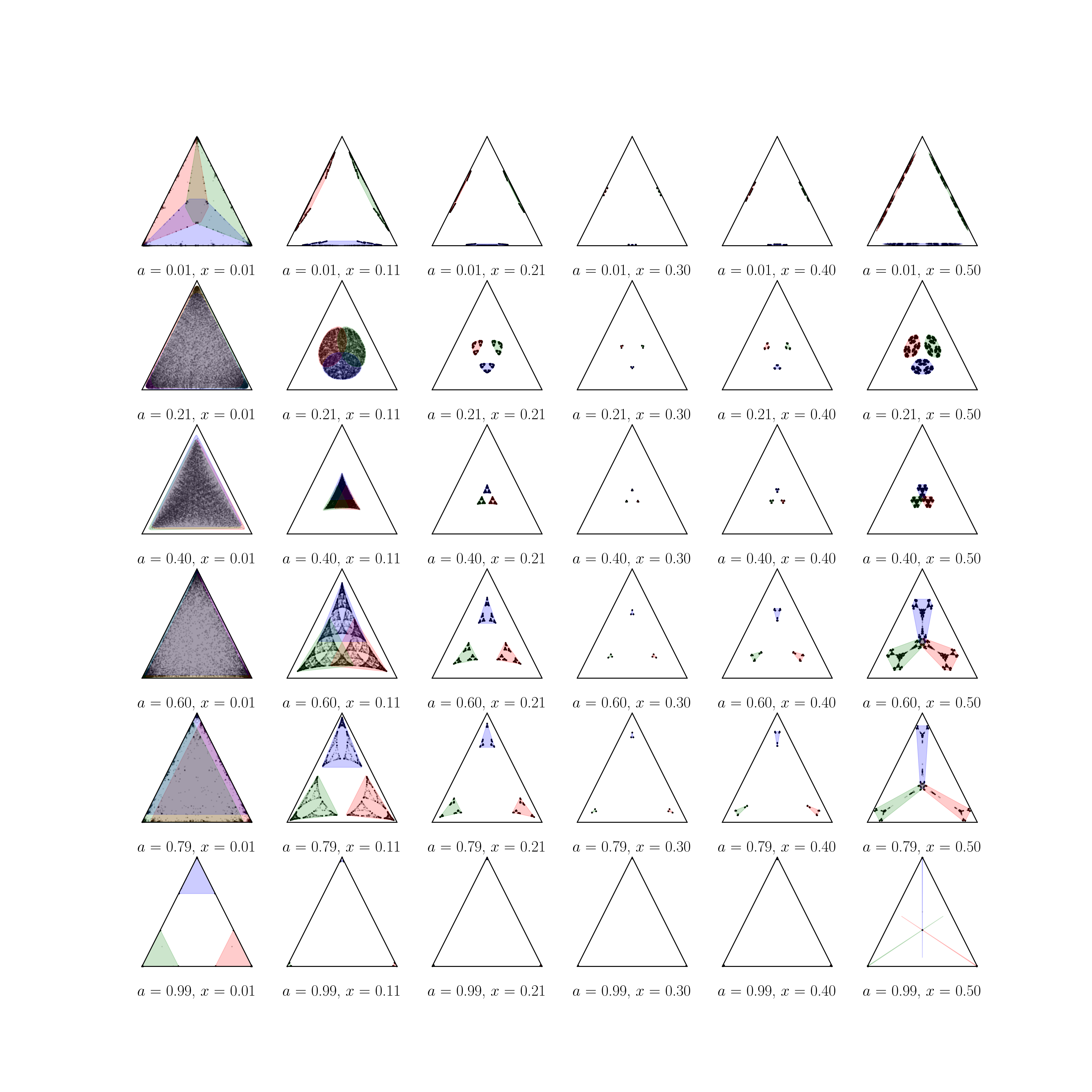}
\caption{Mixed-state attractors generated by a $3$-state HMC parametrized over
	$\alpha \in [0, 1]$ and $x \in [0, 0.5]$. The HMC itself is given in
	\cref{eq:sarah_machine}. $10^5$ mixed states are plotted for each
	attractor, with the initial $5 \times 10^4$ states thrown away as
	transients. The ranges of the symbol-labeled maps are color shaded,
	revealing regions of their image overlap on the attractor. Comparing to
	\cref{eq:sarah_machine}, the red, blue, and green regions represent the
	images of the mapping functions defined by $T^{\square}$, $T^{\triangle}$,
	and $T^{\circ}$, respectively.
	}
\label{fig:sarah_overlap_grid}
\end{figure*}

For two-state nonunifilar HMCs, the mixed-state set lives on the $1$-simplex
$\simplex^1$---the unit interval from $\eta = (0,1)$ to $\eta = (1,0)$. Mixed
states $\mxst \in \MxSSet$ and the dynamic on them exist in a one-dimensional
space and, thus, there is a single negative Lyapunov exponent $\lambda_1 < 0$.

In this case, the calculation of the negative Lyapunov exponent is particularly
direct, since the maps are all one-dimensional. The negative Lyapunov exponent
for a one-dimensional map $\mxst_{n+1} = f(\mxst_n)$ is:
\begin{align*}
\lambda(\mxst_0) = \lim_{N \to \infty} \frac{1}{N} \sum_{i=0}^{N-1} \log 
  \left| \frac{ d f(\mxst_i) }{d \mxst} \right|
\end{align*}
for an orbit starting at $\mxst_0$. For an IFS with a set of of mapping
functions $\{ f^{(\msym)}\}$, we find $\lambda$ as the weighted average of
the Lyapunov exponents of each map: 
\begin{align*}
\lambda_\MxSMeasure = \int \sum_{\msym} p^{(\msym)}(\mxst)
  \log \left| \frac{ d f^{(\msym)} (\mxst) }{d \mxst } \right| d \MxSMeasure
  ~,
\end{align*}
where $\MxSMeasure$ is the IFS's Blackwell measure. We can apply ergodicity to
transform this into a summation over time for ease of calculation.

If a two-state HMC has an MSP with an uncountable infinity of mixed states and
its corresponding IFS satisfies OSC, there is a simple relationship between the
entropy rate, the Lyapunov exponent, and the statistical complexity dimension.
This is given by:
\begin{align}
\dsc (\MxSMeasure)
  = - \frac {h_{\MxSMeasure}} {\lambda_{\MxSMeasure}}
  ~,
\label{eq:IFSInformationDimension}
\end{align}
recalling that $\hmu > 0$, so that the dimension is always positive. Failing
OSC, this ratio is an upper bound on the dimension of the measure $\mu$
\cite{Jaros2008}. For a discussion of the intuition behind this formula, see
\cref{app:BakersMap}.

\begin{figure*}
\captionsetup[subfloat]{width=0.45\textwidth}
\centering
\subfloat[
    Mixed-state attractor area estimated across $\alpha$ and $x$. The area is
    minimized along lines of $\alpha = 1/3$ and $x=1/3$, where the attractor
    becomes point-like. 
    \label{fig:heatmaps_area}]{
    \includegraphics[width=0.46\textwidth]{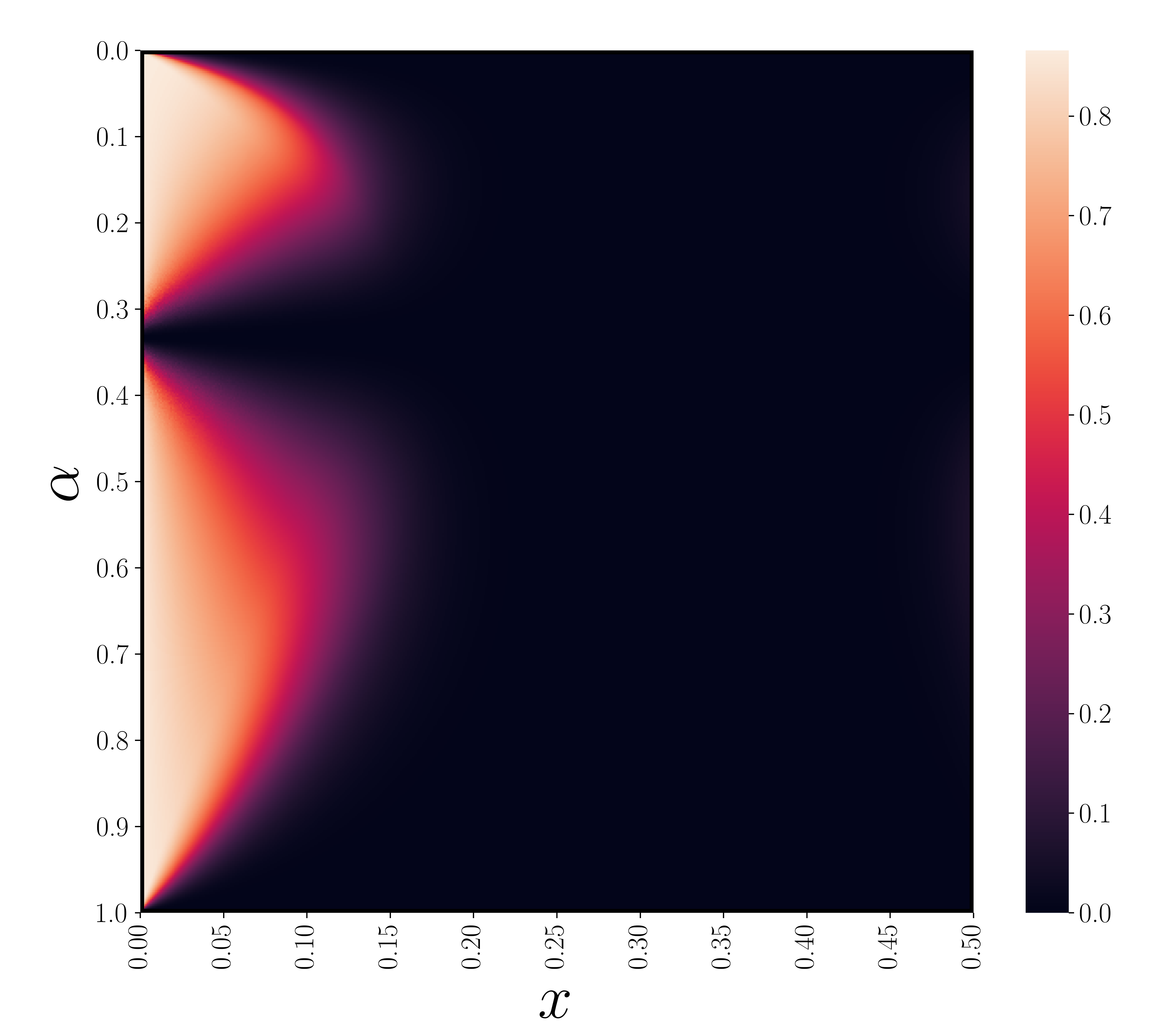}
    } 
\subfloat[
    Mixed-state attractor Lyapunov dimension does not detect the collapse to
    zero dimension along $\alpha=1/3$, which is due to overlaps.
    \label{fig:heatmaps_LEdim}]{
    \includegraphics[width=0.46\textwidth]{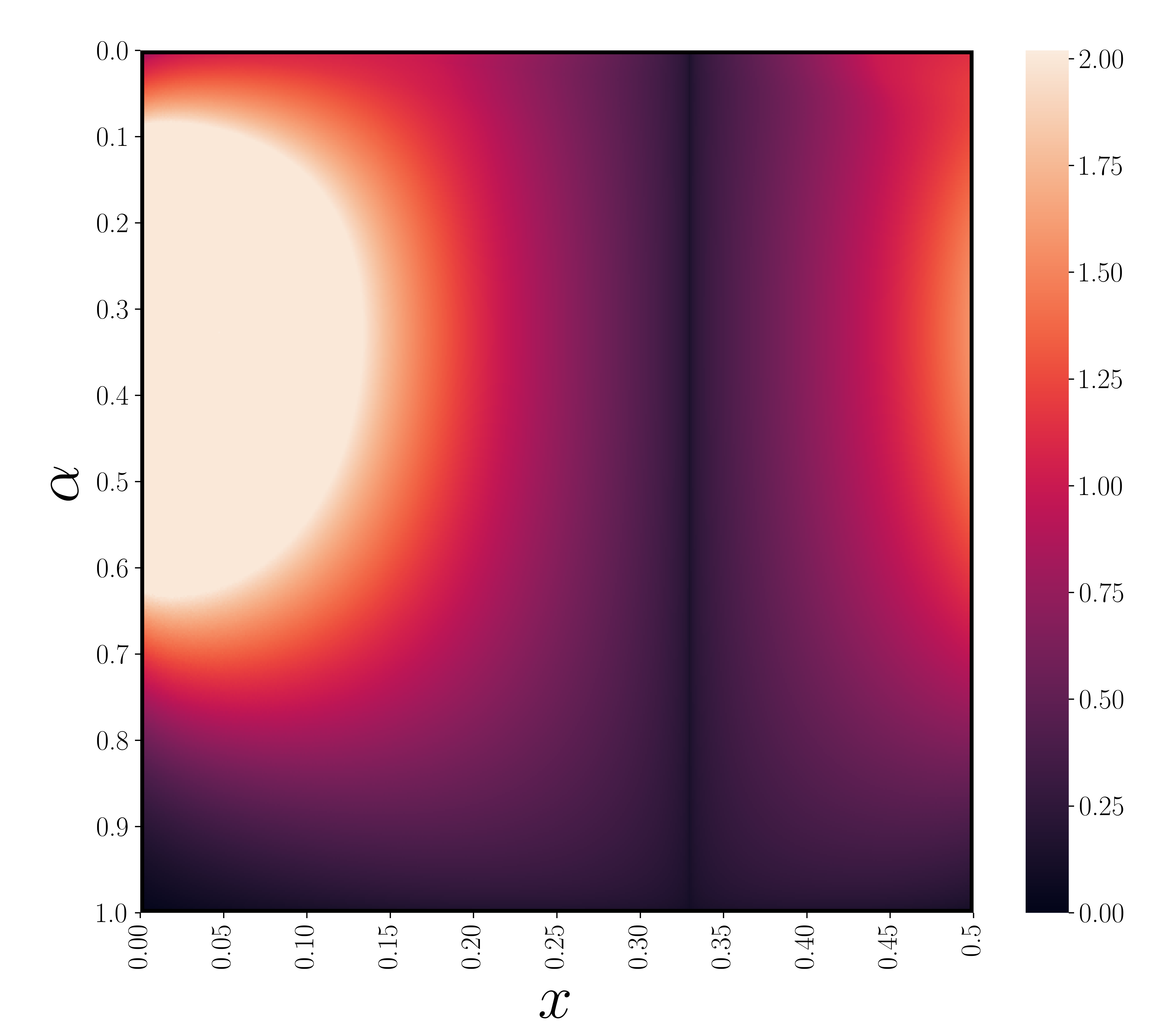}
    }
  \hspace{0mm}
\subfloat[
    Percentage of attractor area in which there is overlap.
    \label{fig:heatmaps_overlaps}]{
    \includegraphics[width=0.46\textwidth]{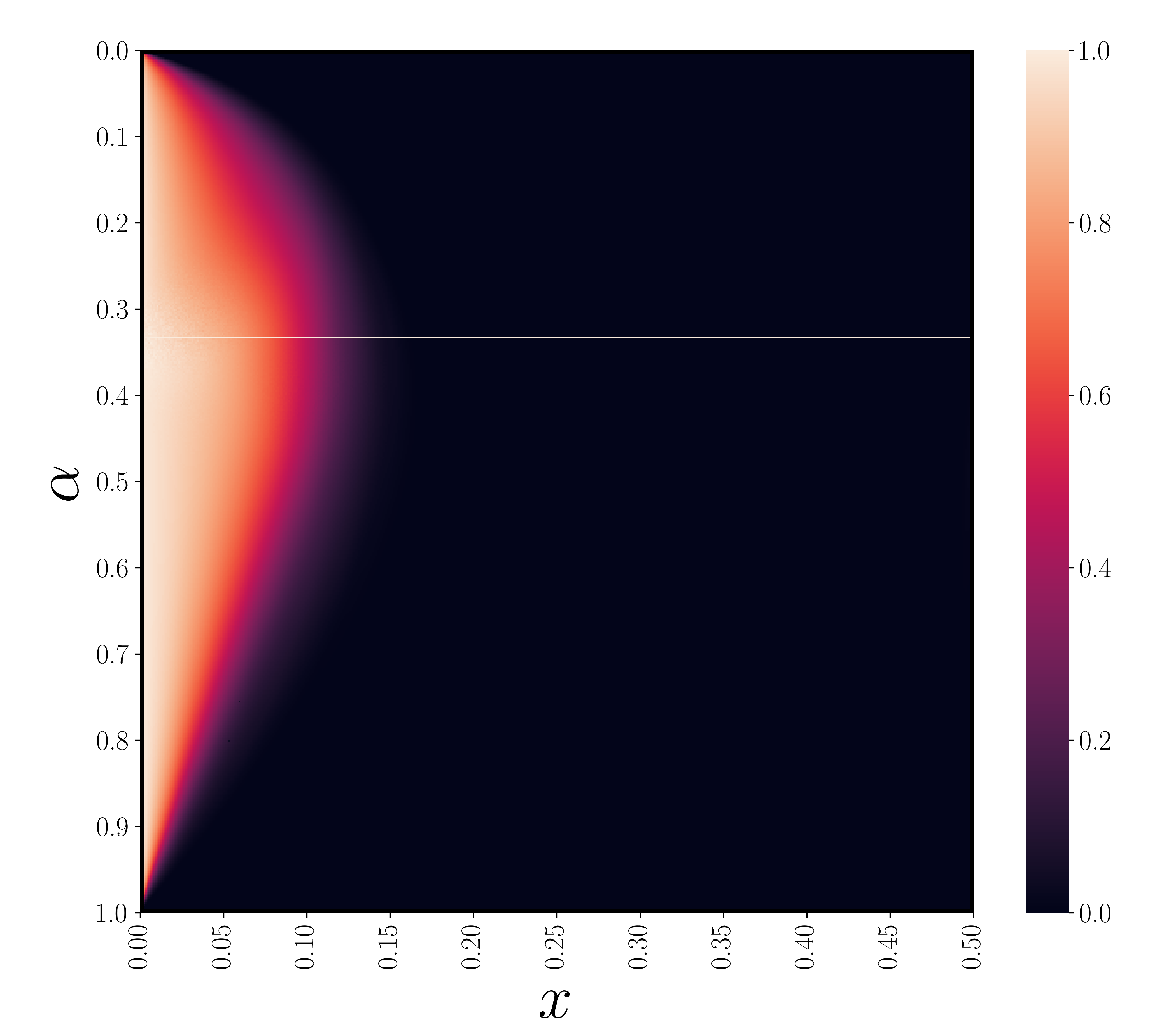}
    }
\subfloat[
    Overlap in the attractor area. Comparing with (c), for much of this area, the
    overlap is very small.
    \label{fig:heatmaps_overlaps_binary}]{
    \includegraphics[width=0.46\textwidth]{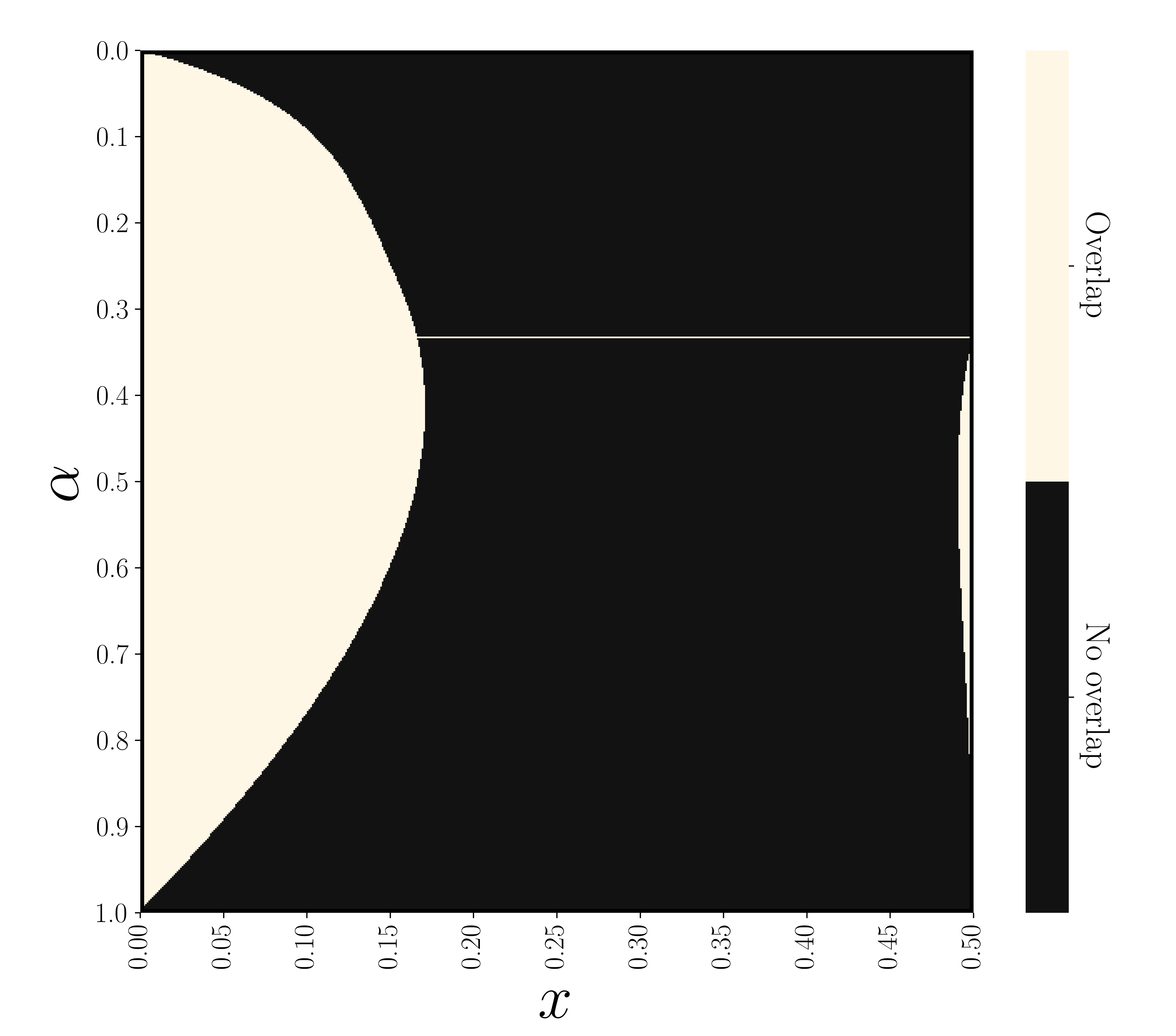}
    } \caption{Attractor area, overlap regions, and Lyapunov dimension of the
mixed-state attractors shown in Fig. \ref{fig:sarah_overlap_grid}, parametrized
by $\alpha = [0, 1]$ and $x = [0,0.5]$. The HMC itself is given in
\cref{sec:Example}. See \cref{app:LyapunovExponents} and
\cref{app:OverlapEstimation} for a detailed discussion of the production of
these plots.} 
\label{fig:heatmaps}
\end{figure*}

\section{Multi-state HMC Examples}
\label{sec:Example}

Notably, the Lyapunov dimension \cref{eq:IFSInformationDimension} for more-than-two-state HMCs is easily shown to be correct when the maps are similitudes and the probability functions are constant. The latter is seen, for example, with the Sierpinski triangle, as discussed in \cref{app:Sierpinski}. 

However, we are generally interested in multi-state HMCs that do not produce
perfectly self-similar fractals. Furthermore, we are often interested in
considering physical systems described by parametrized HMCs, such as those
that arose in the two prequels on quantum measurement processes and information
engine functionality \cite{Vene19a,Jurg20a}. In such cases, an HMC determined
by an application may meet the OSC in some regions of parameter space and fail
to do so in others. Let's consider an HMC that spans the breadth of these
possible behaviors, from zero overlap to complete overlap. This will
demonstrate the range of applicability of our statistical complexity dimension
algorithm. And, ultimately in a sequel \cite{Jurg20e}, it leads to a practical
algorithm and a conjecture for general HMCs.

Consider the following HMC with $3$ symbols and $3$ states:
\begin{align}
\label{eq:sarah_machine}
T^{\square} & = \begin{pmatrix}
        \alpha y & \beta x & \beta x \\
        \alpha x & \beta y & \beta x \\
        \alpha x & \beta x & \beta y
    \end{pmatrix}
    ,~
T^{\triangle} = \begin{pmatrix}
        \beta y & \alpha x & \beta x \\
        \beta x & \alpha y & \beta x \\
        \beta x & \alpha x & \beta y        
    \end{pmatrix}
    ,~\text{and}
	\nonumber
    \\
T^{\circ} & = \begin{pmatrix}
        \beta y & \beta x & \alpha x \\
        \beta x & \beta y & \alpha x \\
        \beta x & \beta x & \alpha y        
    \end{pmatrix}
    ~,
\end{align}
with $\beta = (1-\alpha)/2$ and $y = 1 - 2x$. By inspection, we see that
$\alpha$ takes on any value from $0$ to $1$ and $x$ may range from $0$ to
$1/2$. 

\Cref{fig:sarah_overlap_grid} shows how the MSP attractors change across the
$(\alpha,x)$ parameter space. Each black dot is a generated mixed state, while
the colored regions show the range of each symbol-labeled map.

For example, on one hand, in the top left corner with $\alpha = 0.01$ and
$x = 0.01$, we find an attractor that fills the simplex, with moderate amounts
of overlap. On the other, $\alpha = 0.79$ and $x = 0.11$ produces an attractor
with no overlap, and clearly defined regions.

Moreover, for any $\alpha$, choosing $x=1/3$ leads the MSP attractor to
collapse to a finite $3$-state HMC, since the symbol-labeled mapping functions
become constant functions. In this case, there is no overlap, as each
symbol-labeled map takes on a different constant value.

However, when $\alpha = \beta = 1/3$, all symbol-labeled mapping functions are
identical. Therefore, the attractor is the single fixed-point shared by all
three maps---a single-state HMC. This is a case of maximal possible overlap.
Along both lines in parameter space the MSP collapses to a finite-state HMC, so
$\dsc = 0$, by definition. However, these different mechanisms of
state-collapse are relevant in calculation of $\dLCE$ via
\cref{eq:IFSLyapunovDimension}.

First, consider \cref{fig:heatmaps_area}, which illustrates the estimated area
on the simplex taken up by the attractor across parameter space. For a
discussion of how attractor area was estimated, see
\cref{app:OverlapEstimation}. This figure matches the grid in
\cref{fig:sarah_overlap_grid}: lower values of $x$ produce larger attractors,
excepting the region near $\alpha = 1/3$, where the area drops to zero.

Now, the area taken up by an attractor is not a good proxy for dimension---we
may have very small-in-size attractors with large dimension. Indeed, this is
the case for most values of $\alpha$ near $1/3$, where the attractors are very
small but whose $\dsc \to 2$. However, we know from analyzing the attractor
grid that for HMCs lying exactly on this line, the attractor is instantaneously
finite state. And so, the statistical complexity dimension $\dsc$ must
discontinuously drop to zero. However, this is not accurately reflected by
$\dLCE$, as seen in \cref{fig:heatmaps_LEdim}.

That said, $\dLCE$ clearly smoothly approaches zero as $x \to 1/3$. Along the
vertical line, $\dLCE$ is correctly exactly zero. This is the other line in
parameter space where the MSP is finite state and $\dsc$ is analytically known
to be zero. The disparity, in both correctness and continuity, is due to the
different mechanisms driving the collapse noted above. 

As $x \to 1/3$, the slopes of the symbol-labeled mapping functions approach
zero. When $x = 1/3$, the symbol-labeled mapping functions become constant
values, as reflected in the Lyapunov exponents and consequently in $\dLCE$. The
constant functions have negative Lyapunov exponents of negative infinity,
sending \cref{eq:IFSLyapunovDimension} to zero. 

In contrast, along the $\alpha = 1/3$ line, the contraction in the state space
is a result of the maps instantaneously sharing a fixed point. For $\alpha = 1/3
\pm \epsilon$, the attractor is not finite. In this region of parameter space,
symbol-labeled maps are not infinitely contracting, so $\dLCE$ badly
overestimates $\dsc$ along the $\alpha=1/3$ line. This illustrates the
importance of the OSC on the bound. 

This poses the question, which regions in HMC parameter space exhibit overlap
IFS maps? \Cref{fig:heatmaps_overlaps_binary} depicts parameter space
regions in terms of
overlap or no overlap. \Cref{fig:heatmaps_overlaps} shows this as a percentage
of the total attractor area. For a discussion of how overlap was determined,
please see \cref{app:OverlapEstimation}. Comparing the two, we see that there is
a significant region over which overlap does exist for $x < 0.15$ and a smaller
region where $x > 0.48$. However, for much of that region the attractor's
overlap area is relatively small. As a rule of thumb, the gap between $\dLCE$
and $\dsc$ for the mixed-state set is determined by the percentage of the
attractor that is affected by overlap. If the overlap region is relatively
small, in comparison to the size of the attractor, $\dLCE$ may be very close to
$\dsc$. 

However, if the overlap is very large, $\dLCE$ may be a dramatic overestimation
of $\dsc$. This occurs when $\alpha = 1/3$ and $x < 0.15$. The statistical
complexity dimension $\dsc$ vanishes, yet the Lyapunov dimension saturates at
$\dLCE = 2.0$. 

We also note that the mechanism driving the collapse of the MSP attractor at
$\alpha = 1/3$ is a discontinuity in the parameter space, as compared to $x =
1/3$. This is because state space collapse due to overlap requires the maps to
be identical, and even minute differences in the symbol-labeled transition
matrices will produce an uncountably-infinite MSP, potentially with $\dsc =
2.0$. This encourages us to consider not just the statistical complexity
dimension, but also the area of the attractor and the nearby regions in
parameter space for a clearer understanding of the underlying HMC. In this, the
tools developed here, by allowing (computationally-efficient) surveys of large
regions of parameter space, are particularly useful. 

Thus, while there are wide parameter regions in which the analysis developed
here is correct and efficient, this is not the entire story. Overlap must be
addressed for full generality. Our sequel shows how to correct for
overestimating statistical complexity dimension, allowing accurate calculation
across the entire parameter space \cite{Jurg20e}. However, the exploratory
observations outlined here are a crucial diagnostic guide to those further
developments.

\section{Conclusion}
\label{sec:Conclusion}

Our development opened by considering the challenge of quantifying the
structure of complex systems. For well over a half a century, since Kolmogorov
and Sinai introduced Shannon's mathematical theory of communication to
dynamical systems, the entropy rate stood as the standard by which to quantify
randomness in time series and in chaotic dynamical systems. Quantifying
observable patterns remained a more elusive goal. However, with developments
from computational mechanics, it has become possible to answer questions of
structure and pattern, at least for stochastic processes generated by
finite-state predictive machines, including the symbolic dynamics generated by
chaotic dynamical systems. 

To handle processes generated by finite-state nonpredictive (nonunifilar) hidden
Markov chains, we developed the mixed-state presentation. This
\emph{unifilarizes} general HMCs, giving a predictive presentation that itself
generates the process.  However, adopting a unifilar presentation came at a
heavy cost: Generically, they are infinite state and so previous structural
measures diverge. Nonetheless, we showed how to work constructively with these
infinite mixed-state presentations. In particular, we showed that they fall into
a common class of dynamical system: The mixed-state presentation is a
place-dependent iterated function system.  Due to this, a number of results
from dynamical systems theory can be applied to more fully describe the
original stochastic process.

Previously, others considered the IFS-HMC connection \cite{rezaeian2006hidden,
Slomczynski_2000}. Complementing those efforts, we expanded the role of the
mixed-state presentation to calculate entropy rate and demonstrated its
usefulness in determining the underlying structural properties of the generated
process.  Indeed, \cref{fig:simplex_examples} and \cref{fig:sarah_overlap_grid}
show how visually striking---and distinct---mixed-state sets generated by HMCs
are.

Here, moving in a new direction beyond previous efforts, we established that
the information dimension of the mixed-state attractor is exactly the divergence
rate of the \emph{statistical complexity} \cite{Crut88a}---a measure of a
process' structural complexity that tracks memory. Thus, processes in this
class effectively increase their use of memory, ``creating'' mixed or causal
states, on the fly. Furthermore, we introduced a method to calculate the
information dimension of the mixed-state attractor from the Lyapunov spectrum
of the mixed-state IFS. In this way, we demonstrated that coarse-graining the
mixed-state simplex---the previous method for studying the structure of
infinite-state processes \cite{Marz17a}---can be avoided altogether. This
greatly improves accuracy and computational speed and deepens our understanding
of the origins of complexity in stochastic processes.

During the development, we noted several obstacles. Most importantly, the
presence of \emph{overlap} and failure to meet the Open Set Condition causes
the Lyapunov dimension to be a strict upper bound, and some times quite a poor
one, on the statistical complexity dimension. The final work of our trilogy
\cite{Jurg20e} introduces a measure of how badly the entropy rate overestimates
the expansion of the mixed-state set. Combining this measure with the
Lyapunov-information dimension conjecture finally yields a correct
\cref{eq:IFSLyapunovDimension} to apply to HMCs with overlap; that is, for
processes generated by all general HMCs.

To close, we note that the structural tools introduced here and the
entropy-rate method introduced previously have been put to practical use in two
previous works. One diagnosed the origin of randomness and structural
complexity in quantum measurement \cite{Vene19a}. The other exactly determined
the thermodynamic functioning of Maxwellian information engines \cite{Jurg20a},
when there had been no previous method for this kind of detailed and accurate
analysis. At this point, however, we leave the full explication of these
techniques and further analysis on how mixed states reveal the underlying
structure of processes generated by hidden Markov chains to the sequel
\cite{Jurg20e}. 

\acknowledgements

The authors thank Sam Loomis, Sarah Marzen, Ariadna Venegas-Li, Nicolas Brodu,
Alec Boyd, and Ryan James for helpful discussions and the Telluride Science
Research Center for hospitality during visits and the participants of the
Information Engines Workshops there. JPC acknowledges the kind hospitality of
the Santa Fe Institute, Institute for Advanced Study at the University of
Amsterdam, and California Institute of Technology for their hospitality during
visits. This material is based upon work supported by, or in part by, FQXi
grants FQXi-RFP-IPW-1902 and FQXI-RFP-CPW-2007 and U.S. Army Research Laboratory
and the U.S. Army Research Office grants W911NF-18-1-0028 and W911NF-21-1-0048.

\appendix

\onecolumngrid
\clearpage
\begin{center}
{\huge Supplementary Materials}\\
\vspace{0.1in}
{\LARGE \ourTitle}\\[15pt]
{\large Alexandra Jurgens and James P. Crutchfield\\[5pt]
\arxiv{\arXivID}
}
\end{center}

\setcounter{equation}{0}
\setcounter{figure}{0}
\setcounter{table}{0}
\setcounter{page}{1}
\setcounter{section}{0}
\makeatletter
\renewcommand{\theequation}{S\arabic{equation}}
\renewcommand{\thefigure}{S\arabic{figure}}
\renewcommand{\thetable}{S\arabic{table}}

The Supplementary Materials give the HMCs for the example processes considered,
review MSP minimality, draw a correspondence with the Baker's Map on the unit
square, layout an HMC whose mixed-state set $\MxSSet$ is the well-known
Sierpinski Triangle fractal, and review Lyapunov characteristic exponents and
calculating their spectrum for IFSs.

\section{Nonunifilar HMC Examples}
\label{app:nonunifilarHMCs}

We reproduce here the HMCs used to create \cref{fig:simplex_examples}. First,
the ``alpha'' HMC, from \cref{fig:simplex_example_alpha}, is given by::

\begin{align}
  \label{eq:alpha_machine}
  T^{\square} & = \begin{pmatrix}
    2.734 \times 10^{-2} & 0.392 & 1.924 \times 10^{-2} \\
    0.475 & 2.176 \times 10^{-2} & 2.766 \times 10^{-4} \\
    0.224 & 2.711 \times 10^{-3} & 0.236
      \end{pmatrix}
      , \\
  T^{\triangle} & = \begin{pmatrix}
    1.845 \times 10^{-3} & 0.133 & 0.259 \\
    3.913 \times 10^{-2} & 0.315 & 2.789 \times 10^{-2} \\
    0.467 & 1.015 \times 10^{-2}  & 4.699 \times 10^{-3}        
      \end{pmatrix}
      ,~\text{and}~
    \nonumber
      \\
  T^{\circ} & = \begin{pmatrix}
    9.782 \times 10^{-2} & 3.374 \times 10^{-2} & 3.644 \times 10^{-2}\\
    5.422 \times 10^{-2} & 6.503 \times 10^{-2} &  2.090 \times 10^{-3} \\
    5.328 \times 10^{-2} & 1.278 \times 10^{-3} & 8.778 \times 10^{-4}       
      \end{pmatrix}
      ~,
    \nonumber
\end{align}

\cref{fig:simplex_example_sarah} is given by \cref{eq:sarah_machine}, at $\alpha
= 0.6$ and $x = 0.1$. The ``beta'' HMC, in \cref{fig:simplex_example_beta},
is given by:

\begin{align}
  \label{eq:beta_machine}
  T^{\square} & = \begin{pmatrix}
    5.001 \times 10^{-2}  & 0.388 & 4.251 \times 10^{-2} \\
    0.464 & 4.484 \times 10^{-2} & 2.495 \times 10^{-2} \\
    0.232  & 2.720 \times 10^{-2} & 0.243
      \end{pmatrix}
      , \\
  T^{\triangle} & = \begin{pmatrix}
    1.708 \times 10^{-3} & 0.123 & 0.240 \\
    3.623 \times 10^{-2} & 0.292  & 2.583 \times 10^{-2} \\
    0.432  & 9.397 \times 10^{-3} & 4.351 \times 10^{-3}      
      \end{pmatrix}
      ,~\text{and}~
    \nonumber
      \\
  T^{\circ} & = \begin{pmatrix}
    9.0576 \times 10^{-2}  & 3.124 \times 10^{-2}  & 3.374 \times 10^{-2} \\
    5.020 \times 10^{-2} & 6.021 \times 10^{-2} & 1.935 \times 10^{-3}\\
    4.933 \times 10^{-2} & 1.183 \times 10^{-3} & 8.127 \times 10^{-4}      
      \end{pmatrix}
      ~,
    \nonumber
\end{align}

Due to finite numerical accuracy, reproduction of the attractors using these
specifications may differ slightly from \cref{fig:simplex_examples}. 

\section{Mixed-State Presentation Minimality}
\label{app:MinimalityofMSP}

Given an HMC $M$, minimality of infinite-state mixed-state presentations
$\MSP(M)$ is an open question. MSPs are not guaranteed to be minimal. In fact,
it is possible to construct MSPs with an uncountably-infinite number of states
for a process that requires only one state to optimally predict, as seen with
the Cantor machine process in \cref{Fig:cantormachine,Fig:mixedstatemethod}.
Note that while this HMC generates an uncountable number of mixed states, each
one has the same emitted-symbol probability distribution, indicating that all
states can be merged into a single state with no loss of predictability.
Indeed, the \eM for the HMC depicted in \cref{Fig:cantormachine} is simply the
single-state Fair Coin HMC. 

A proposed solution to this needless presentation verbosity is a short and
simple check on \emph{mergeablility} of mixed states. This refers to any two
distinct mixed states that have the same conditional probability distribution
over future strings; i.e., any two mixed states $\mxst_0$ and $\mxstalt_0$ for
which:
\begin{align}
\Pr( \MS{0}{L} | \mxst_0 ) = \Pr( \MS{0}{L} | \mxstalt_0 )
  ~,
\label{eq:MergeableStates}
\end{align}
for all $L \in \mathbb{N}^+$.

\begin{figure}
  \centering
  \includegraphics{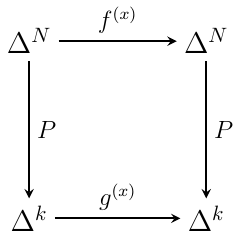}
  \caption[text]{Commuting diagram for probability functions $P = \{p^{(x)} \}$,
    mixed-state mapping functions $f^{(x)}$, and proposed
    symbol-distribution mapping functions $g^{(x)}$.
    }
  \label{Fig:simplexdiagram}
\end{figure}

A benefit of the IFS formalization of the MSP is the ability to directly check
for duplicated states and therefore determine if the MSP is nonminimal. We check
this by considering, for an $N+1$ state HMC $M$ with alphabet $\MeasAlphabet
= \{ 0, 1, \dots, k \}$, the dynamic not only over mixed states, but probability
distributions over symbols. Let:
\begin{align}
P(\mxst) = \left( p^{(0)}(\mxst), \dots, p^{(k-1)}(\mxst) \right)
\label{eq:VectorProbabilityFunction}
\end{align}
and consider \cref{Fig:simplexdiagram}. For each mixed state $\mxst \in
\simplex^N$, \cref{eq:VectorProbabilityFunction} gives the corresponding
probability distribution $\rho(\mxst) \in \simplex^k$ over the symbols $\msym
\in \MeasAlphabet$. Let $M$ emit symbol $\msym$, then the dynamic from one such
probability distribution $\rho \in \simplex^{k}$ to the next is given by:
\begin{align}
g^{(\msym)}(\rho_t) & = P \circ f^{(\msym)} \circ P^{-1} (\rho) \nonumber \\
  & = \rho_{t+1, \msym}
  ~.
\label{eq:ProabilityVectorDynamic}
\end{align}

From this, we see that if \cref{eq:ProabilityVectorDynamic} is invertible,
$g^{(\msym)} : \simplex^k \to \simplex^k$ is well defined and has the same
functional properties as $f^{(\msym)}$. In other words, in this case, it is not
possible to have two distinct mixed states $\mxst, \mxstalt \in \simplex^N$ with
the same probability distribution over symbols. And, the probability
distributions can only converge under the action of $g^{(\msym)}$ if the mixed
states also converge under the action of $f^{(\msym)}$. When every mixed state
has a unique outgoing probability distribution, these states are also the causal
states, and the MSP is the process' \eM. Our companion work \cite{Jurg20d}
elaborates on this and the implications for identifying the embedding dimension
of minimal generators.

\section{Correspondence with Baker's Map}
\label{app:BakersMap}

The simple dimension formula in \cref{eq:IFSInformationDimension} may not seem
easily motivated. Especially, considering that, in general, both positive and
negative Lyapunov exponents are required to have a nontrivial attractor.
However, for iterated function systems, all Lyapunov exponents are negative and
the expansive role played by positive Lyapunov exponents is instead played by
an IFS's stochastic map selection, as measured by the entropy rate $\hmu$.

This is more intuitively appreciated by comparing the two-state IFS with the
Baker's map. Consider the Baker's map:
\begin{align*}
x_{n+1} & = \left\{\begin{aligned}
    &\frac{x_n}{s_0}  ~, & y < p\\
    &\frac{x_n+s_1-1}{s_1}  ~, & y \geq p
    \end{aligned}
    \right.
	\text{~~~and}\\
y_{n+1} & = \left\{
        \begin{aligned}
            &\frac{y_n}{p}  ~, & y < p\\
            &\frac{y_n - p}{p - 1}  ~, & y \geq p
        \end{aligned}
    \right.
\end{align*}
It has LCE spectrum $\Lambda = \{\lambda_1,\lambda_2\}$, where:
\begin{align*}
\lambda_1 & = p \log(p) + (1-p) \log (1-p) \\
\lambda_2 & = p \log(1/s_0) + (1-p) \log (1/s_1)
  ~.
\end{align*}
Note that $\lambda_1 > 0$ and $\lambda_2 < 0$. Then, the Lyapunov dimension is:
\begin{align*}
\dLCE = 1 - \frac{\lambda_1}{\lambda_2}
  ~.
\end{align*}

To compare this to an IFS, take:
\begin{align*}
\{f(x)\} & = \left\{\frac{x_n}{s_0}, \frac{x_n+s_1-1}{s_1}\right\} ~\text{and}\\
\{p(x)\} & = \{p, 1-p\}
  ~.
\end{align*}
Thus, we identify the
$y$ coordinate as controlling the stochastic map choice. The dynamic over
position in the $y$ direction exactly determines the IFS entropy rate. Since the
Baker's map is volume preserving in $y$, the extra dimension always contributes
a plus one in the dimension formula. In other words, the dimension along a slice
of constant $y$ equals the IFS dimension.

\section{Sierpinski's Triangle}
\label{app:Sierpinski}

The Sierpinski triangle is a canonical Cantor set in two dimensions. An HMC that
generates a MSP attractor that is the Sierpinski triangle is:
\begin{align}
\label{eq:sierpinski}
T^{(0)} & = \begin{pmatrix}
                a & 0 & a(s-1) \\
                0 & a & a(s-1) \\
                0 & 0 & as
             \end{pmatrix}
	,~ 
T^{(1)} = \begin{pmatrix}
                \frac{1-as}{2} & 0 & 0 \\
                \frac{(1-as)(s-1)}{2s} & \frac{1-as}{2s} & 0 \\
                \frac{(1-as)(s-1)}{2s} & 0 & \frac{1-as}{2s}
             \end{pmatrix}
	,~\text{and}~
T^{(2)} = \begin{pmatrix}
               \frac{1-as}{2s} & \frac{(1-as)(s-1)}{2s} & 0 \\
               0 & \frac{1-as}{2} & 0 \\
               0 & \frac{(1-as)(s-1)}{2s} & \frac{(1-as)(s-1)}{2s}
            \end{pmatrix}
  ~,
\end{align}
where $s$ controls the contraction coefficient and $a$ controls the probability
of selecting the maps. This HMC produces constant probability functions: 
\begin{align*}
p^{(0)} & = as ~,~
p^{(1)} = \frac{1 - as}{2} ~,~\text{and}~
p^{(2)} = \frac{1-as}{2}
  ~.
\end{align*}
and, therefore, linear mappings, since $f^{(0)} = \bra{\mxst} T^{(i)} /
p^{(i)}(\mxst)$. The constant probability functions make the entropy rate
trivial to calculate. And, the linearity of the mappings does the same for the
Lyapunov exponents.

\begin{figure*}
\centering
\includegraphics[width=.7\textwidth]{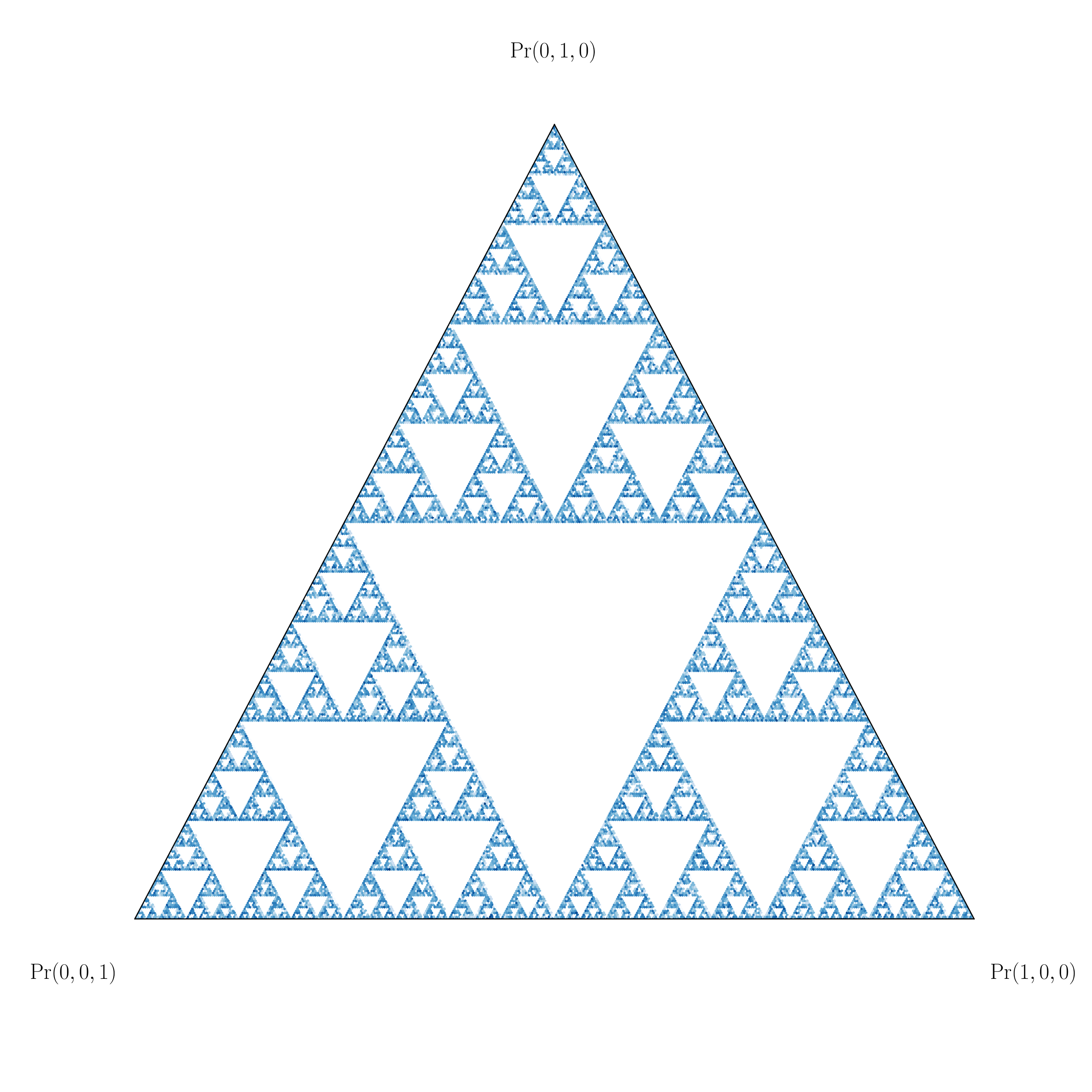}
\caption{Attractor of the $3$-state, $3$-symbol machine specified in
	\cref{eq:sierpinski}, with $s = 2$ and $a = \frac{1}{6}$.
	}
\label{fig:sierpinski}
\end{figure*}

Setting $s = 2$ and $a = 1 / 6$, results in equal probability for all maps and
gives the standard Sierpinski triangle shown in \cref{fig:sierpinski}. In this
case, the entropy rate is $\hmu = \log_2 3$ and the Lyapunov exponents are
both $-\log_2 2$. Plugging this into \cref{eq:IFSLyapunovDimension} returns
the well-known fractal dimension of the Sierpinski triangle: $\log_2 3 / \log_2
2 \approx 1.585$. 

\section{Lyapunov Exponents}
\label{app:LyapunovExponents}

A \emph{Lyapunov characteristic exponent} for a dynamical system measures the
exponential rate of separation of trajectories that begin infinitesimally close.
Since, typically, the separation rate depends on the direction of the initial
separation, we use a spectrum of Lyapunov exponents, with one exponent for each
state-space dimension. In a chaotic dynamical system, at least one Lyapunov
exponent is positive. In general, the Lyapunov exponent spectrum for an
$N$-dimensional dynamical system with mapping $x_{n+1} = F(x_n)$ depends on the
initial condition $x_0$. However, here we consider ergodic systems, for which
the spectrum does not.

Consider the map's \emph{Jacobian} matrix:
\begin{align*}
J = \frac{\partial F}{\partial x}
\end{align*}
and the evolution of vectors in the tangent space, controlled by:
\begin{align*}
\dot{Y} = YJ ~,
\end{align*}
where $Y(0) = \mathbb{I}_N$ and $Y(t)$ describes how an infinitesimal change in
$x(0)$ has propagated to $x(t)$. Let $\{y_1, \dots, y_N\}$ be the eigenvalues of
the matrix $Y(t)Y(t)^\intercal$. Then, the Lyapunov exponents are:
\begin{align*}
\lambda_i = \lim_{t\to \infty} \frac{1}{2t} \log y_i
  ~.
\end{align*}
The \emph{Lyapunov numbers} were introduced and proven to exist by Oseledets
\cite{Oseledets68}. The Lyapunov exponents are merely the logarithms of Lyapunov
numbers.

The most common way of calculating an IFS's Lyapunov spectrum, employed to
produce the results in \cref{fig:heatmaps}, is the \emph{pull-back} method. The
basic idea is that for an IFS in the $N-1$ simplex, defined by an $N$-state HMC,
there will be $N-1$ independent directions of contraction. These directions are
represented by a coordinate frame of $N-1$ vectors that are kept orthogonal and
normalized. This coordinate frame is carried along a long orbit of the IFS. At
each time step, the Jacobian is used to evolve the frame. We track the
contraction rate of each vector, and this becomes the estimation of the Lyapunov
exponents. 

In contrast to applying this approach to deterministic dynamical systems, as is
more familiar, an IFS's stochastic nature introduces additional error. Since the
Jacobian varies not just across the simplex, but also for the selected maps, the
orbit must be long enough to sample from the IFS attractor's distribution
accurately for each possible mapping function. This being said, it has
been established that the pull-back method works for IFS spectra, given a
sufficiently long orbit \cite{Elton1987}.

The prequel, on estimating the entropy rate of HMC processes, made use of
error-bounding techniques from Markov chain Monte Carlo (MCMC) \cite{Jurg20b}.
Since here we are estimating Lyapunov exponents by sampling from the Blackwell
distribution, similar error-bounding techniques apply. In this analysis, there
are two fundamental sources of estimation error. First, that due to
\emph{initialization bias} or undesired statistical trends introduced by the
initial transient data produced by the Markov chain before it reaches the
desired stationary distribution. Second, there are errors induced by
\emph{autocorrelation in equilibrium}. That is, the samples produced by the
Markov chain are correlated. And, the consequence is that statistical error
cannot be estimated by $1 / \sqrt{N}$, as done for $N$ independent samples.

Bounding these error sources requires estimating the autocorrelation function,
which can be done from long sequences of samples. If we have the nonunifilar
model in hand, it is a simple matter of sweeping through increasingly long
sequences of generated samples until we observe convergence of the
autocorrelation function. An alternative method of approximating the
infinite-state HMC with a finite-state approximation is discussed in detail in
our previous work \cite{Jurg20b}. The upshot is that the method here generally
efficiently leads to accurate estimates of the LCE spectrum.

For completeness, we note that there are alternative methods to calculate
Lyapunov exponents; see, e.g., Refs. \cite{Froy00a, Boyarsky92}. These methods
may be more appropriate in specific applications. That said, the accuracy and
applicability of Lyapunov exponent estimation is not the focus here.

\section{Overlap Estimation}
\label{app:OverlapEstimation}

To estimate the size of a mixed-state attractor and overlap of mapping functions
in \cref{fig:heatmaps}, a combination of techniques were used. We will briefly
summarize the method here.

First, $250,000$ different HMCs were generated using a $500 \times 500$
parameter grid over $\alpha = [0, 1]$ and $x = [0,0.5]$. Each HMC was defined by
plugging the appropriate parameter values into the symbol-labeled transition
matrices in \cref{eq:sarah_machine}. From this HMC, the mapping and probability
functions were defined (see \cref{eq:MxStUpdate,eq:SymbolFromMixedState}), producing a place-dependent IFS. 

For each IFS and associated HMC, $10,000$ mixed states were generated from an
initial randomized state, throwing away the first $5,000$ as transients. Using a
spatial algorithm from the SciPy Python package, a convex hull was drawn around
this set of points, with a small buffer. This convex hull (the attractor
``outline'') was converted into a polygon. This polygon was then evolved
independently by each symbol-labeled mapping function, producing three polygons,
each associated with a symbol. This may be visualized by referencing
\cref{fig:sarah_overlap_grid}, where the evolved polygons are depicted on top of
the mixed-state attractor, each with a different color. We can see that the
combination of these polygons must necessarily cover the attractor. 

These three symbol-labeled polygons were then combined into a single polygon or
multipolygon (a polygon with ``holes'' that are themselves polygons) using the
geometry-processing module Shapely \cite{shapely}. This produces a more accurate
outline of the attractor than the convex hull. This process may then be repeated
with the new outline for as many iterations as desired, until a polygon or
multipolygon that covers the mixed state attractor with the desired level of
accuracy is produced. The same result could be achieved by beginning with the
entire simplex as the initial outline, without any production of mixed-states.
However, the step of estimating the convex hull sharply reduces the number of
required iterations and, more importantly, makes the required number more equal
across parameter space. To see this, consider that attractors taking up less
of the simplex require several iterations to converge to the small size of
the attractor. By initializing with the convex hull, the process of converging
to the attractor's basic shape is skipped, and the iterations are merely
refinements.

In producing \cref{fig:heatmaps}, we found that we could determine good outlines
across parameter space by evolving the convex hull three times. To produce
\cref{fig:heatmaps_area}, the area of the resultant polygon or multipolygon was
found using Shapely. To produce \cref{fig:heatmaps_overlaps_binary} and
\cref{fig:heatmaps_overlaps_binary}, the outline was evolved one more time by
each map, and the resultant polygons and/or multipolygons were checked for
intersection. For the binary overlap/no-overlap plot in
\cref{fig:heatmaps_overlaps_binary}, only the existence of overlap somewhere on
the attractor was considered. For the percentage overlap in
\cref{fig:heatmaps_overlaps}, the area of the total outline that was comprised
of overlapping polygons---whether only two or all three---was compared to the
total area. The subtlety of whether an overlap region included two or three
maps was largely ignored here, but will be analyzed in future explorations of
the overlap problem.

\end{document}